\documentclass[12pt]{article}
\usepackage{amsmath}
\usepackage{amsthm}
\usepackage{amsfonts}
\usepackage{amssymb}
\usepackage{fullpage}
\usepackage{mathtools}
\usepackage{graphicx}
\usepackage{color}
\usepackage{times}
\usepackage{subcaption}
\usepackage{verbatim}

\begin{document}
\renewcommand{\thefootnote}{\fnsymbol{footnote}}
\title{On Pareto Joint Inversion of guided waves}
\author{
Adrian Bogacz\footnote{Department of Geoinformatics and Applied Computer Science, AGH University of Science and Technology, Krak\'ow, Poland, {\tt abogacz@agh.edu.pl}},
David R. Dalton\footnote{
Department of Earth Sciences, Memorial University of Newfoundland,
{\tt dalton.nfld@gmail.com}}, 
Tomasz Danek\footnote{Department of Geoinformatics and Applied Computer Science, AGH University of Science and Technology, Krak\'ow, Poland, {\tt tdanek@agh.edu.pl}},
Katarzyna Miernik\footnote{Department of Geoinformatics and Applied Computer Science, AGH University of Science and Technology, Krak\'ow, Poland, {\tt kmiernik@agh.edu.pl}},\\
Michael A. Slawinski\footnote{
Department of Earth Sciences, Memorial University of Newfoundland,
{\tt mslawins@mac.com}}}
\date{\today}
\maketitle
\renewcommand{\thefootnote}{\arabic{footnote}}
\setcounter{footnote}{0}
\section*{Abstract}
We use the Pareto Joint Inversion, together with the Particle Swarm Optimization, to invert the Love and quasi-Rayleigh surface-wave speeds, obtained from dispersion curves, in order to infer the elasticity parameters, mass densities and layer thickness of the model for which these curves are generated.
For both waves, we use the dispersion relations derived by Dalton et al.~\cite{dalton}.
Numerical results are presented for three angular frequencies, $15$\,, $60$ and $100~{\rm s}^{-1}$\,, and for two, five and seven modes, respectively. 
Comparisons of the model parameters with the values inverted with error-free input indicate an accurate process. 
If, however, we introduce a $5\%$ error to the input, the results become significantly less accurate, which indicates that the inverse operation, even though stable, is error-sensitive.
Correlations between the inverted elasticity parameters indicate that the layer parameters are more sensitive to input errors than the halfspace parameters.
In agreement with Dalton et al.~\cite{dalton}, the fundamental mode is mainly sensitive to the layer parameters whereas higher modes are sensitive to both the layer and halfspace properties; for the second mode, the results are more accurate for low frequencies.
\section{Introduction}
Two types of guided waves can propagate in an elastic layer overlying an elastic halfspace (e.g.,~Dalton et al.~\cite{dalton}).
On the surface, their displacements are perpendicular to the direction of propagation.
Though the displacements for both waves are parallel to the surface, the displacement of one of them is perpendicular to the direction of propagation and, for the other, parallel to that direction.
The former is called the Love wave, and the latter the quasi-Rayleigh wave, where the prefix distinguishes it from the Rayleigh wave that exists in the halfspace alone and, in contrast to its guided counterpart, is not dispersive.
The orthogonal polarization of the displacement vectors on the surface allows us to identify each wave and to distinguish between them.
Hence, we can use their propagation speeds on the surface to infer information about the model in which they propagate.
To do so, we use the Pareto Joint Inversion.
\section{Previous work}
There are several important contributions to inversion of dispersion relations of guided waves for model parameters.
Let us comment on ones with a particular relevance to our work.

A common technique to obtain quasi-Rayleigh-wave dispersion curves is the Multichannel Analysis of Surface Waves technique (Park et al.,~\cite{park}). 
An approach to inverting such curves for multiple layers is given in Xia et al.~\cite{xia}, who use the Levenberg-Marquardt and singular-value decomposition techniques to analyze the Jacobian matrix, and demonstrate sensitivity of material properties to the dispersion curve.

Wathelet et al.~\cite{wathelet} use a neighbourhood algorithm, which is a stochastic direct-search technique, to invert quasi-Rayleigh-wave dispersion curves obtained from ambient vibration measurements.
Lu et al.~\cite{lu} invert quasi-Rayleigh waves in the presence of a low-velocity layer, using a genetic algorithm.
Boxberger et al.~\cite{boxberger} perform a joint inversion, based on a genetic algorithm, using quasi-Rayleigh and Love wave dispersion curves and Horizontal-to-Vertical Spectral Ratio curves obtained from seismic noise array measurements.
Fang et al.~\cite{fang} invert dispersion data without generating phase or group velocity maps, using raytracing and a tomographic inversion.
Xie and Liu~\cite{xie} do Love-wave inversion for a near-surface transversely isotropic structure, using the Very Fast Simulated Annealing algorithm.

Wang et al.~\cite{wang} use phase velocity inversion, based on first-order perturbation theory, including multiple modes and both quasi-Rayleigh and Love waves, to examine intrinsic versus extrinsic radial anisotropy in the Earth; the latter anisotropy refers to a homogenized model.  
Wang et al.~\cite{wang} use the classical iterative quasi-Newton method to minimize the $L_2$ norm misfit and introduce the Generalized Minimal Residual Method.

Dal Moro and Ferigo~\cite{dalmoro2011} carry out a Pareto Joint Inversion of synthetic quasi-Rayleigh and Love-wave dispersion curves for a multiple-layer model using an evolutionary algorithm optimization scheme.   
Dal Moro~\cite{dalmoro2010} examines a Pareto Joint Inversion using an evolutionary algorithm of the combined quasi-Rayleigh and Love wave dispersion curves and Horizontal-to-Vertical Spectral Ratio data.
Dal Moro et al.~\cite{dalmoro2015} perform a three-target Pareto Joint Inversion based on full velocity spectra, using an evolutionary algorithm optimization scheme.
\section{Dispersion relations}
To derive dispersion relations, Dalton et al.~\cite{dalton} consider an elastic layer of thickness~$Z$ overlying an elastic halfspace.
Using Cartesian coordinates, we set the surface at~$x_3=0$\,, and the interface at~$x_3=Z$\,, with the $x_3$-axis positive downward.
The layer consists of mass density,~$\rho^u$\,, and elasticity parameters, $C_{11}^u$ and $C_{44}^u$\,.
The quantities of the halfspace are denoted with superscript~$d$\,.
The same quantities can be expressed in terms of the $P$ and $S$ wave speeds,
\begin{equation*}
\alpha^{(\cdot)}=\sqrt{\frac{C_{11}^{(\cdot)}}{\rho^{(\cdot)}}}\,,\qquad\beta^{(\cdot)}=\sqrt{\frac{C_{44}^{(\cdot)}}{\rho^{(\cdot)}}}\,.	
\end{equation*}
For the Love wave, the dispersion relation is
\begin{equation}
\label{lovedet}
\mathrm{D}_\ell(v_\ell)=\det
\left[\begin{array}{ccc}
{\rm e}^{\iota b'_\ell}&-{\rm e}^{-\iota b'_\ell}&0\\
-\iota s^u_\ell C_{44}^u&\iota s^u_\ell C_{44}^u&s^d_\ell C_{44}^d\\
1&1&-1
\end{array}\right]
=2 s^u_\ell C_{44}^u\sin b'_\ell-2 s^d_\ell C_{44}^d \cos b'_\ell=0\,,
\end{equation}
where
\begin{equation*}
\kappa_\ell=\omega/v_\ell\,,\quad s^u_\ell=\sqrt{\frac{v_\ell^2}{(\beta^u)^2}-1}\,,\quad s^d_\ell=\sqrt{1-\frac{v_\ell^2}{(\beta^d)^2}}\,,\quad
b_\ell'=\kappa_\ell s^u_\ell Z\,.
\end{equation*}
This equation has real solutions,~$v_\ell$\,, for $\beta^u < v_\ell < \beta^d$\,, which are referred to as modes; each solution can be represented by a dispersion curve of~$v_\ell$ plotted against $\omega$\,, along which $\mathrm{D}_\ell$ is zero.
The solution with the lowest value of $v_\ell$ for a given $\omega$ is called the fundamental mode.

Formally, the dispersion relation for the quasi-Rayleigh wave is given in terms of the determinant of a~$6\times 6$ matrix, which,
as shown by Dalton et al.~\cite{dalton}, can be reduced to the determinant of a~$2\times 2$ matrix:

\begin{equation}
\label{eq:ourdet}
\mathrm{D}_r(v_r)=
4\,C^u_{44} \det\left[
\begin{array}{cc} 
s^u X & s^u S\\r^u T & r^u Y
\end{array}
\right]=4\,C^u_{44} r^u s^u(XY-ST)=4\,C^u_{44} r^u s^uD_{r2}\,,
\end{equation}
where
\begin{equation*}
X :=
\left[(s^u)^2)-1\right]\left[-(v_{r}^2 q+ 2 p) B' +2 p r^d \cos b'\right] +
2 \left[r^u ( 2 p -v_{r}^2 \rho^d)\sin a' + r^d( 2 p + v_{r}^2 \rho^u)\cos a'\right],
\end{equation*}
\begin{equation*}
Y := \left[(s^u)^2)-1\right]\left[(v_{r}^2 q+ 2 p) A' -2 p s^d \cos a'\right] +
2 \left[- s^d ( 2 p +v_{r}^2 \rho^u)\cos b' - s^u( 2 p - v_{r}^2 \rho^d)\sin b'\right],
\end{equation*}
\begin{equation*}
S := \left[(s^u)^2)-1\right]\left[-(v_{r}^2\rho^u+ 2 p)s^d B' + (2 p-v_{r}^2 \rho^d) \cos b'\right] +
2 \left[ ( 2 p +v_{r}^2 q)\cos a' + 2 p r^u s^d \sin a'\right],
\end{equation*}
\begin{equation*}
T := \left[(s^u)^2)-1\right]\left[r^d (v_{r}^2 \rho^u + 2 p) A' -(2 p-v_{r}^2 \rho^d) \cos a'\right] - 
2 \left[( 2 p +v_{r}^2 q)\cos b' +2  s^u r^d p \sin b'\right],
\end{equation*}
with~$\kappa,q, p, A', B',r^u,s^u,r^d,s^d,a',b',a,b$ given by
$\kappa=\omega/v_r
$\,,
$
q:=\rho^u-\rho^d
$\,,
$
p:=C^d_{44}-C^u_{44}
$\,,
\begin{equation*}
A':=
\begin{dcases}
\frac{\sin a'}{r^u}&r^u\neq0\\
\kappa Z&r^u=0
\end{dcases}\,,
\qquad
B':=
\begin{dcases}
\frac{\sin b'}{s^u}&s^u\neq0\\
\kappa Z&s^u=0
\end{dcases}\,,
\end{equation*}
\begin{equation*}
r^u=\sqrt{\frac{v_r^2}{(\alpha^u)^2}-1}\,,\quad
s^u=\sqrt{\frac{v_r^2}{(\beta^u)^2}-1}\,,\quad
r^d=\sqrt{1-\frac{v_r^2}{(\alpha^d)^2}}\,,\quad
s^d=\sqrt{1-\frac{v_r^2}{(\beta^d)^2}}
\end{equation*}
and
\begin{equation*}
a'=\kappa r^u Z\,, \quad b'=\kappa s^u Z\,, \quad  a=\kappa r^d Z\,, \quad b=\kappa s^d Z\,.
\end{equation*}
These equations include several corrections to the formulas on p.~200 of Ud\'{\i}as~\cite{udias} (Dalton et al.~\cite{dalton}).
Values of $\mathrm{D}_r$ can be imaginary if the product of $r^u$ and $s^u$ is imaginary.

For modes other than the fundamental mode, equation~(\ref{eq:ourdet}) has a solution only for $\beta^u<v_r<\beta^d<\alpha^d$
and usually, but not always, for $v_r > \alpha^u$ (Ud\'{\i}as,~\cite{udias}). 
For the fundamental mode there is a solution for $v_r<\beta^u$\,, for higher values of $\omega$\,.
For $v_r>\alpha^u$\,, the determinant is real;
for $\beta^u<v_r<\alpha^u$\,, the determinant is imaginary; for $v_r<\beta^u$ the determinant is real.
\section{Pareto Joint Inversion}
\label{sec:pareto}
In this study, the dispersion relations of the Love and quasi-Rayleigh waves are the two target functions to be examined together by the Pareto Joint Inversion.
In general, we search for
\begin{equation}
\label{eq:min}
{\rm min}[f_{1}(x),f_{2}(x),\,\dots\,,f_{n}(x)]\,,\qquad x \in S\,,
\end{equation}
where $f_i$ are target functions and $S$ is the space of acceptable solutions.
In this study, $f_1=\mathrm{D}_{r2}$ and $f_2=\mathrm{D}_\ell$\,, given, respectively, in expression~(\ref{lovedet}) and expression~(\ref{eq:ourdet}), above.
Every solution is
\begin{equation}
\label{eq:set}
C=\{ y \in{\mathbb R}^{n}: y=f(x) : x \in S\}\,.
\end{equation}
Among them, a Pareto solution is vector~$x^*\!\in S$\,, such that all other vectors of this type return a higher value of at least one of the functions, $f_i$\,.

The set of all Pareto optimal solutions is~${\cal P}^*$ and
${\cal PF}^*=(f_{1}(x),f_{2}(x),\dots,f_{n}(x))$\,, where $x\in{\cal P}^*$\,, is the Pareto front.
Each iteration of the algorithm generates a single Pareto solution to be added to the Pareto front.
In this paper, the Particle Swarm Optimization algorithm  (Kennedy and Eberhart~\cite{kennedy},  Parsopoulos and Vrahatis~\cite{parsopoulos}) is used to obtain each element of the Pareto front.

The target function for Love waves is equation \eqref{lovedet}.
Since this is a purely real equation, computations involving imaginary numbers are not necessary.

We use $D_{r2}=XY-ST$ as the target function for quasi-Rayleigh waves; it is real for the input values of $C_{11}^u$\,, $C_{44}^u$\,, $\rho^u$\,, $C_{11}^d$\,, $C_{44}^d$\,, $\rho^d$ and $Z$\,.
Even in this case, since the parameters do not have any constraints in the calculations, which involve randomization, complex numbers still appear.
However, using $\sin(\iota x)=\iota\sinh(x)$ and $\cos(\iota x)=\cosh(x)$, we restrict our computations to real numbers.
\section{Numerical results and discussion}
\subsection{Input}
\begin{figure}
\centerline{
\includegraphics[scale=0.55]{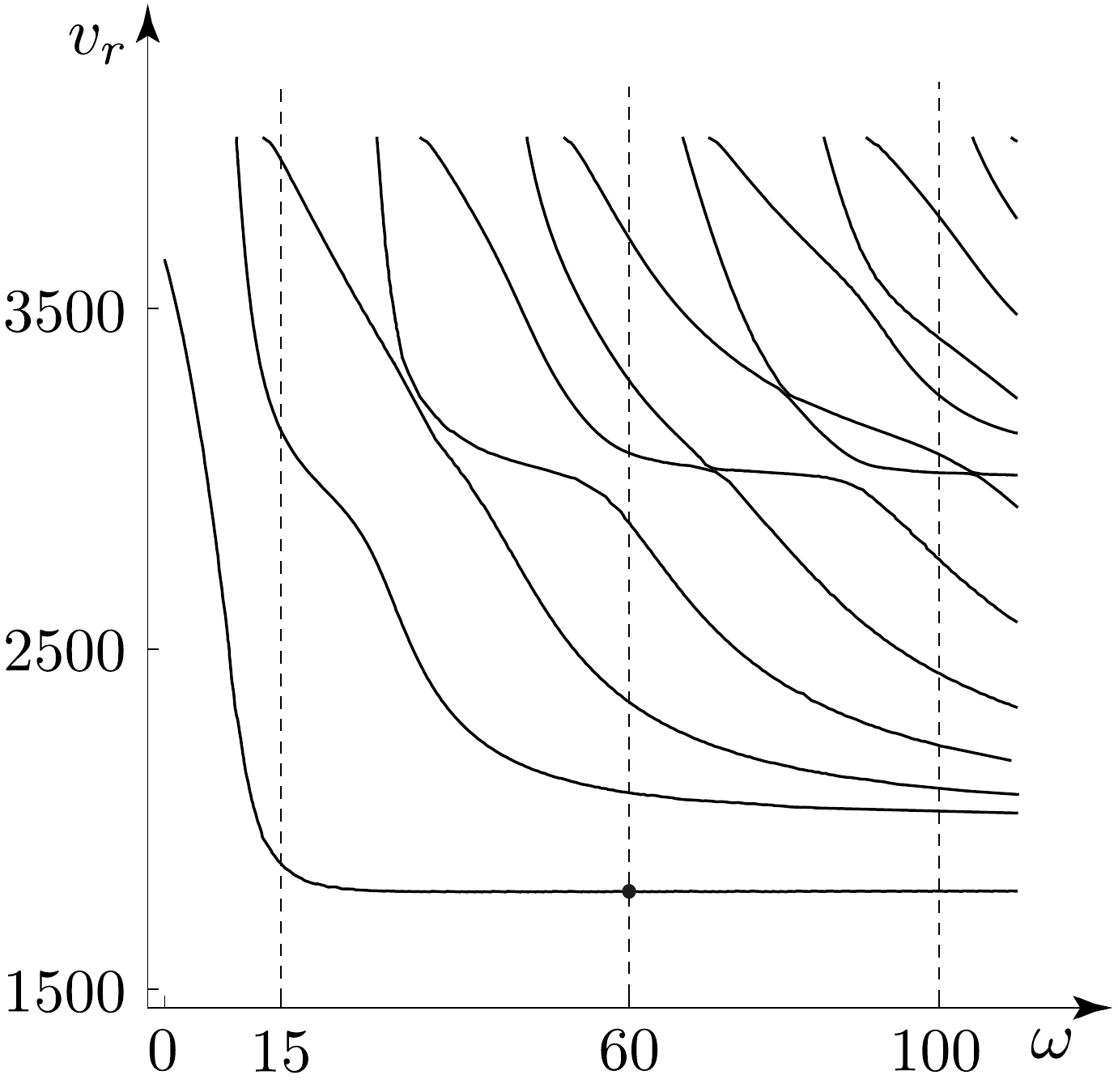}
\hfill
\includegraphics[scale=0.55]{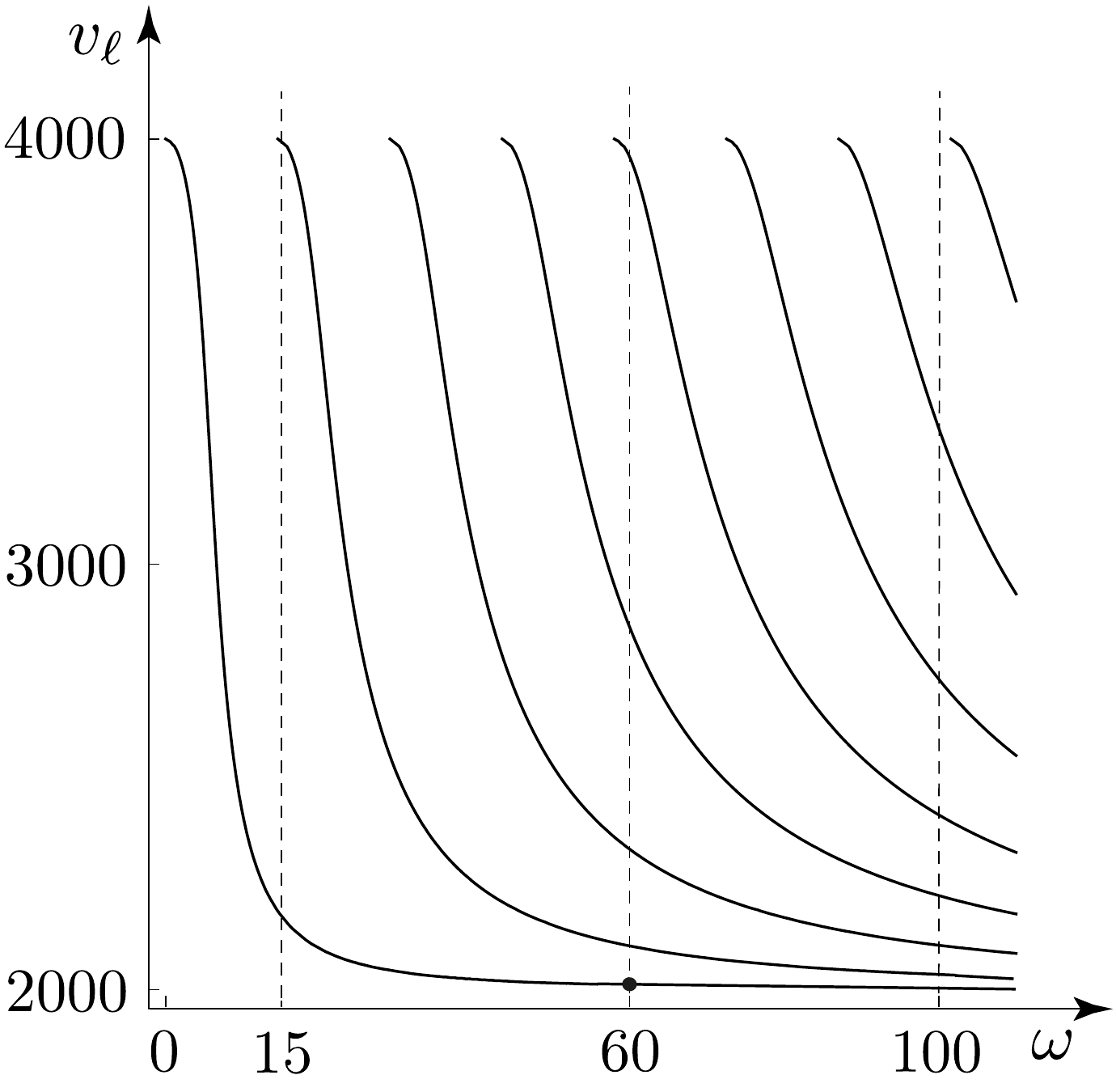}
}
\caption{Dispersion curves of quasi-Rayleigh wave, in the left-hand plot, and of Love wave, in the right-hand plot}
\label{fig:dispersion}
\end{figure}

Figure~\ref{fig:dispersion} illustrates the dispersion curves for the quasi-Rayleigh and Love waves, which are used as input data for the inversion.
Along each curve, the corresponding determinantal equation is satisfied.
Each curve corresponds to a single mode, with the lowest curve being the fundamental mode.
The dashed lines correspond to the angular frequencies for which we perform the inversion.
Their intercepts with dispersion curves correspond to propagation speeds---along the surface---for distinct modes of a guided wave.
We use the first two modes for $\omega=15~{\rm s}^{-1}$, the first five modes for $\omega=60~{\rm s}^{-1}$, and the first seven modes for $\omega=100~{\rm s}^{-1}$.
The dots, in Figure~\ref{fig:dispersion}, correspond to the error-free values of~$v_r$ and $v_\ell$\,, used to obtain the histograms in Figures~\ref{fig:h_1}--\ref{fig:h_3}; the values used to obtain histograms in Figures~\ref{fig:h_4}--\ref{fig:h_6} are $1.05\,v_r$ and $1.05\,v_\ell$\,.

\subsection{Input without errors}
\begin{figure} 
\begin{subfigure}{0.45\textwidth}
\includegraphics[width=\linewidth]{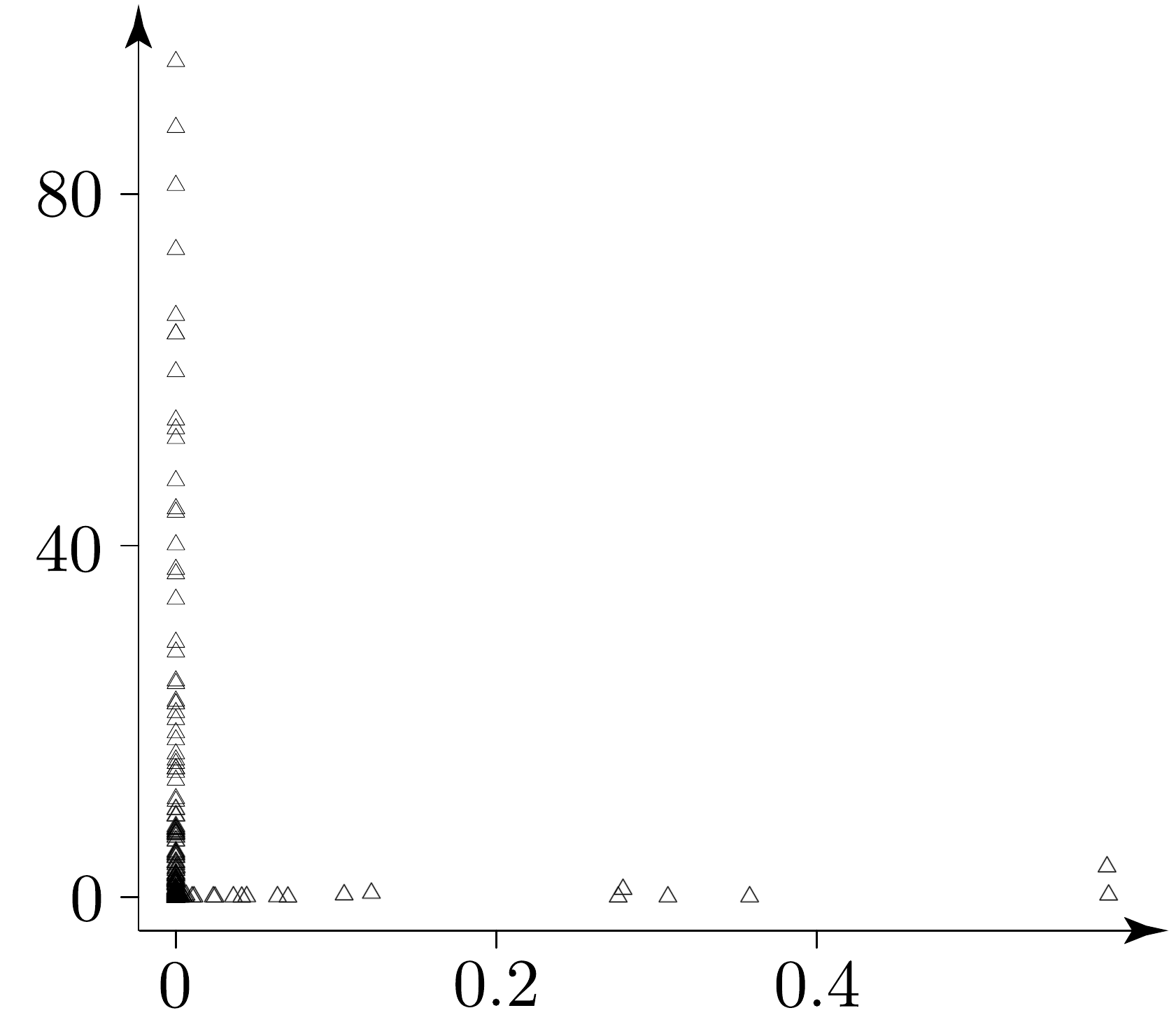}
\caption{}\label{fig:Pareto_15Hz_m1}
\end{subfigure}
\hspace*{\fill} 
\begin{subfigure}{0.45\textwidth}
\includegraphics[width=\linewidth]{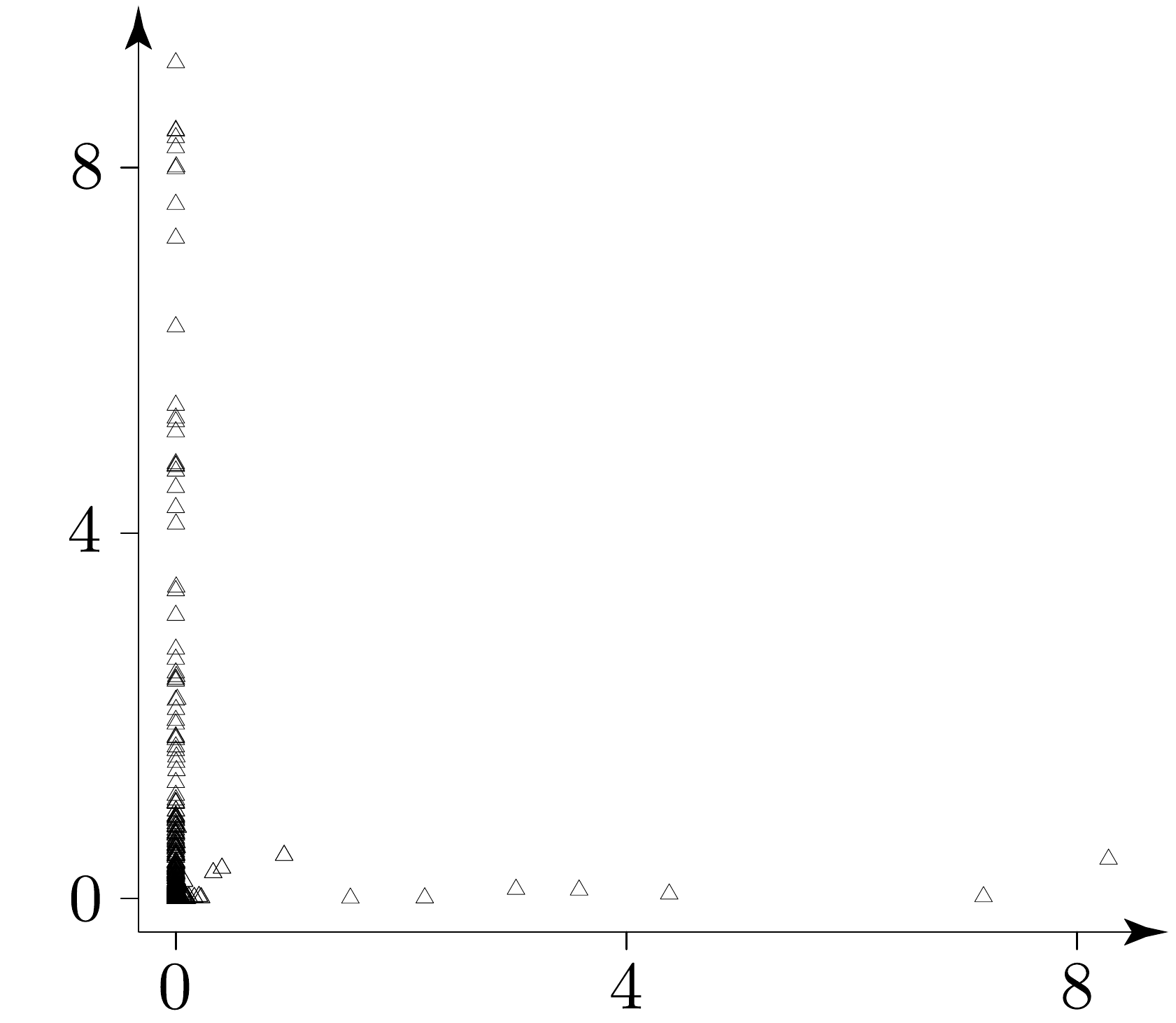}
\caption{}\label{fig:Pareto_15Hz_m2}
\end{subfigure}
\hspace*{\fill} 

\caption{Pareto fronts for $\omega=15~{\rm s}^{-1}$; plots~(a) and (b): first mode and second mode}
\label{fig:Pareto_15Hz_testII}
\end{figure}


\begin{figure} 
\hspace*{\fill} 
\begin{subfigure}{0.4\textwidth}
\includegraphics[width=\linewidth]{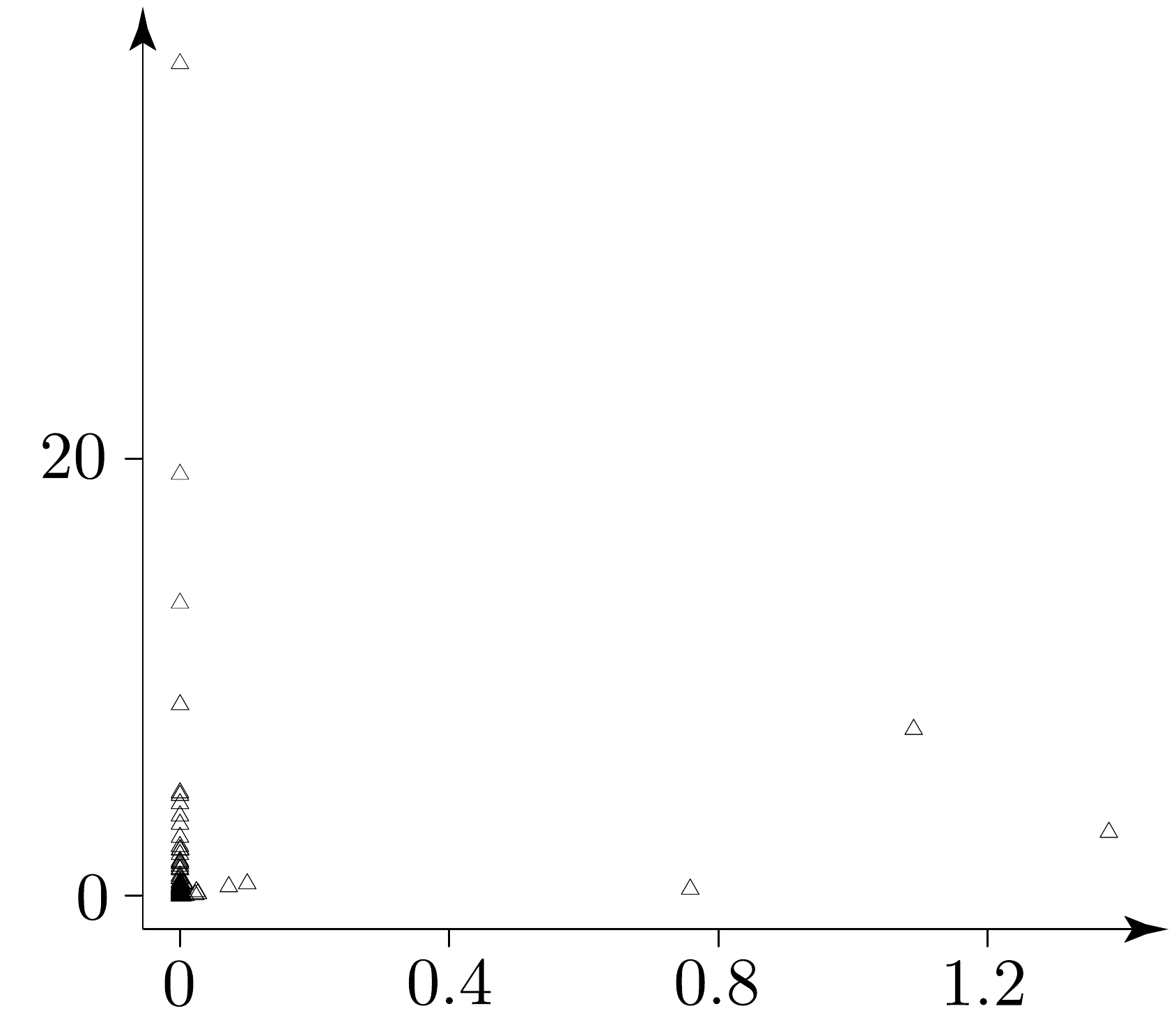}
\caption{}\label{fig:Pareto_60Hz_m1}
\end{subfigure}
\hspace*{\fill} 
\begin{subfigure}{0.4\textwidth}
\includegraphics[width=\linewidth]{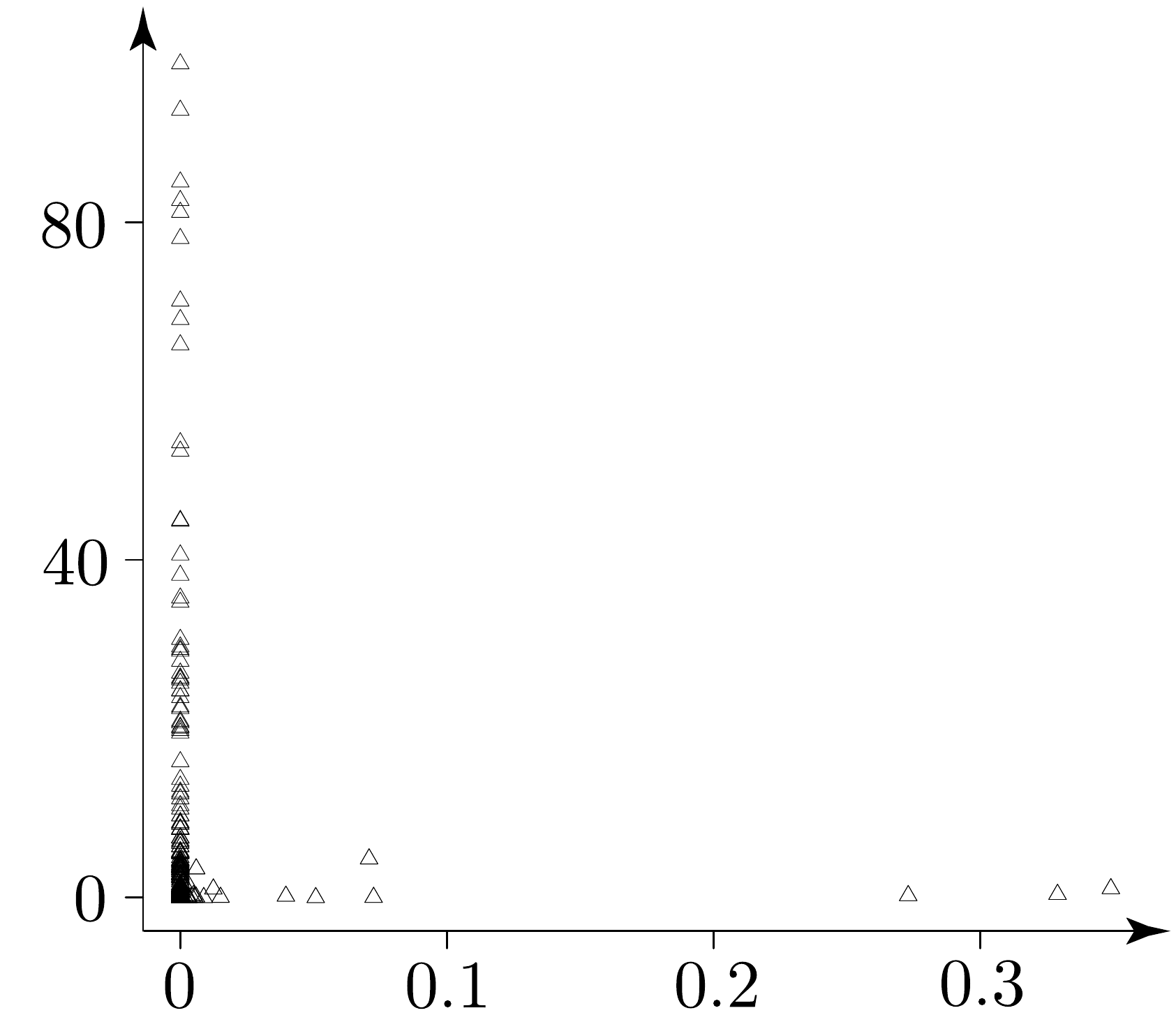}
\caption{}\label{fig:Pareto_60Hz_m2}
\end{subfigure}
\hspace*{\fill} 
\begin{subfigure}{0.4\textwidth}
\includegraphics[width=\linewidth]{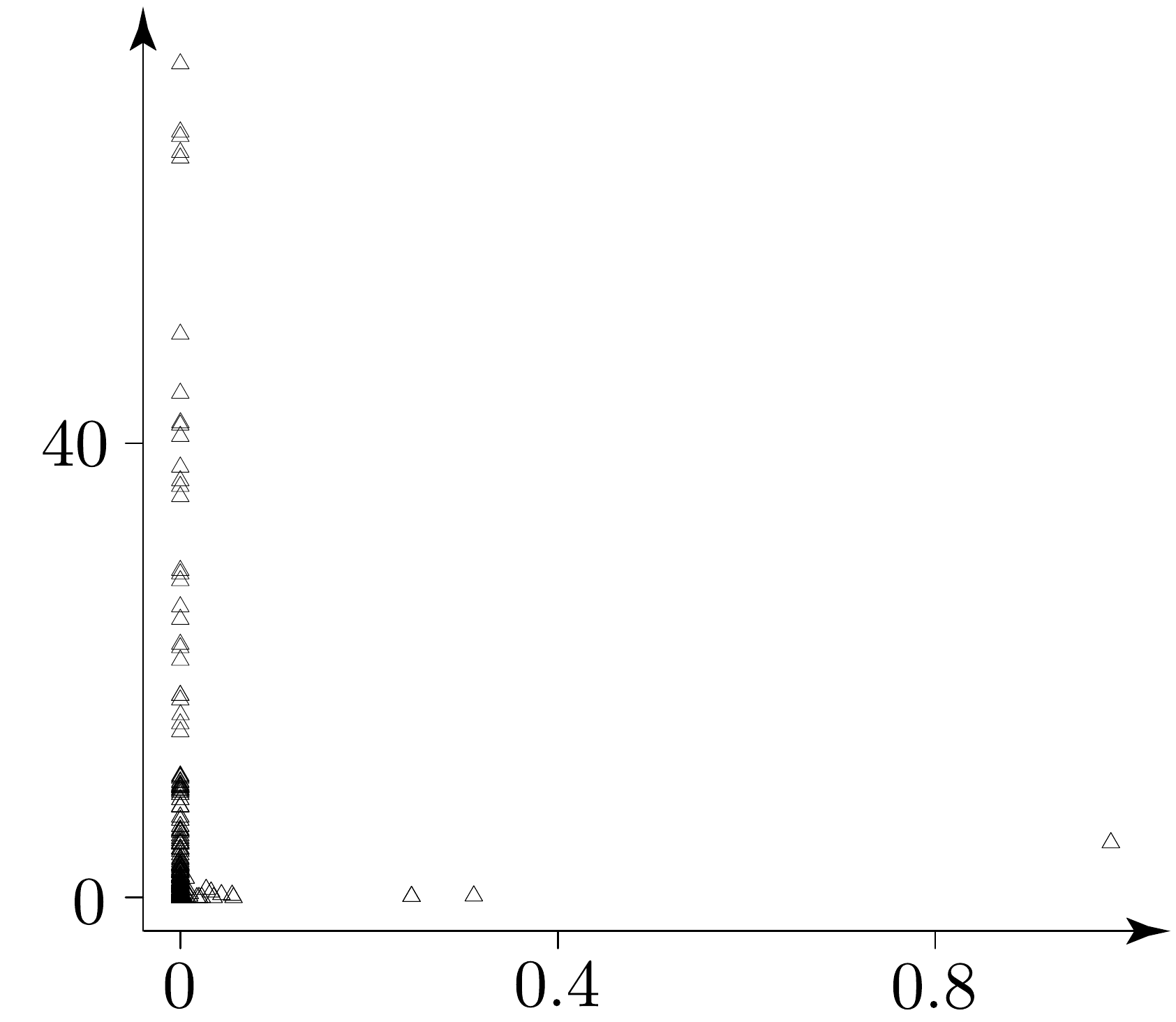}
\caption{}\label{fig:Pareto_60Hz_m3}
\end{subfigure}
\hspace*{\fill} 
\begin{subfigure}{0.4\textwidth}
\includegraphics[width=\linewidth]{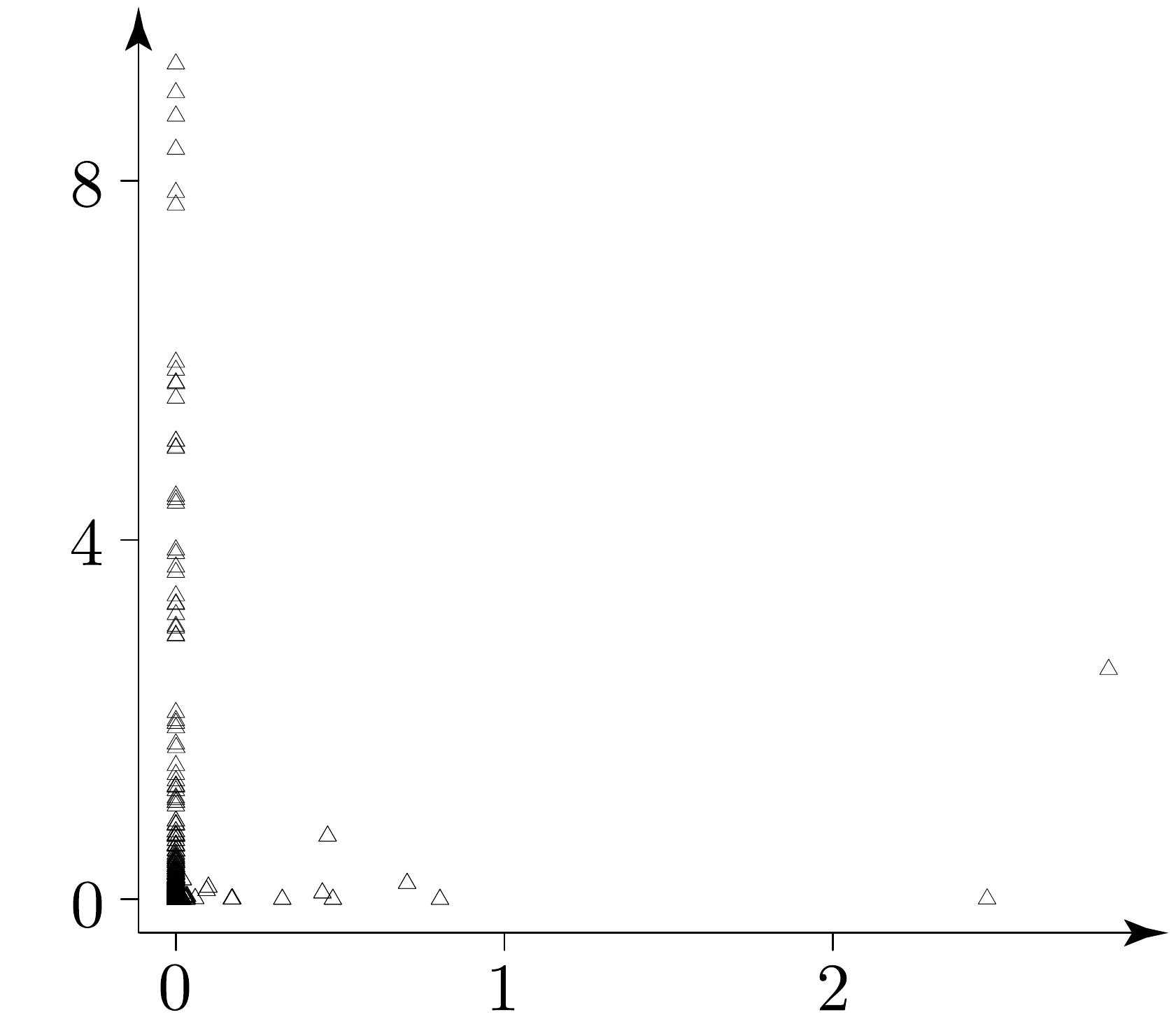}
\caption{}\label{fig:Pareto_60Hz_m4}
\end{subfigure}
\hspace*{\fill} 
\begin{subfigure}{0.4\textwidth}
\includegraphics[width=\linewidth]{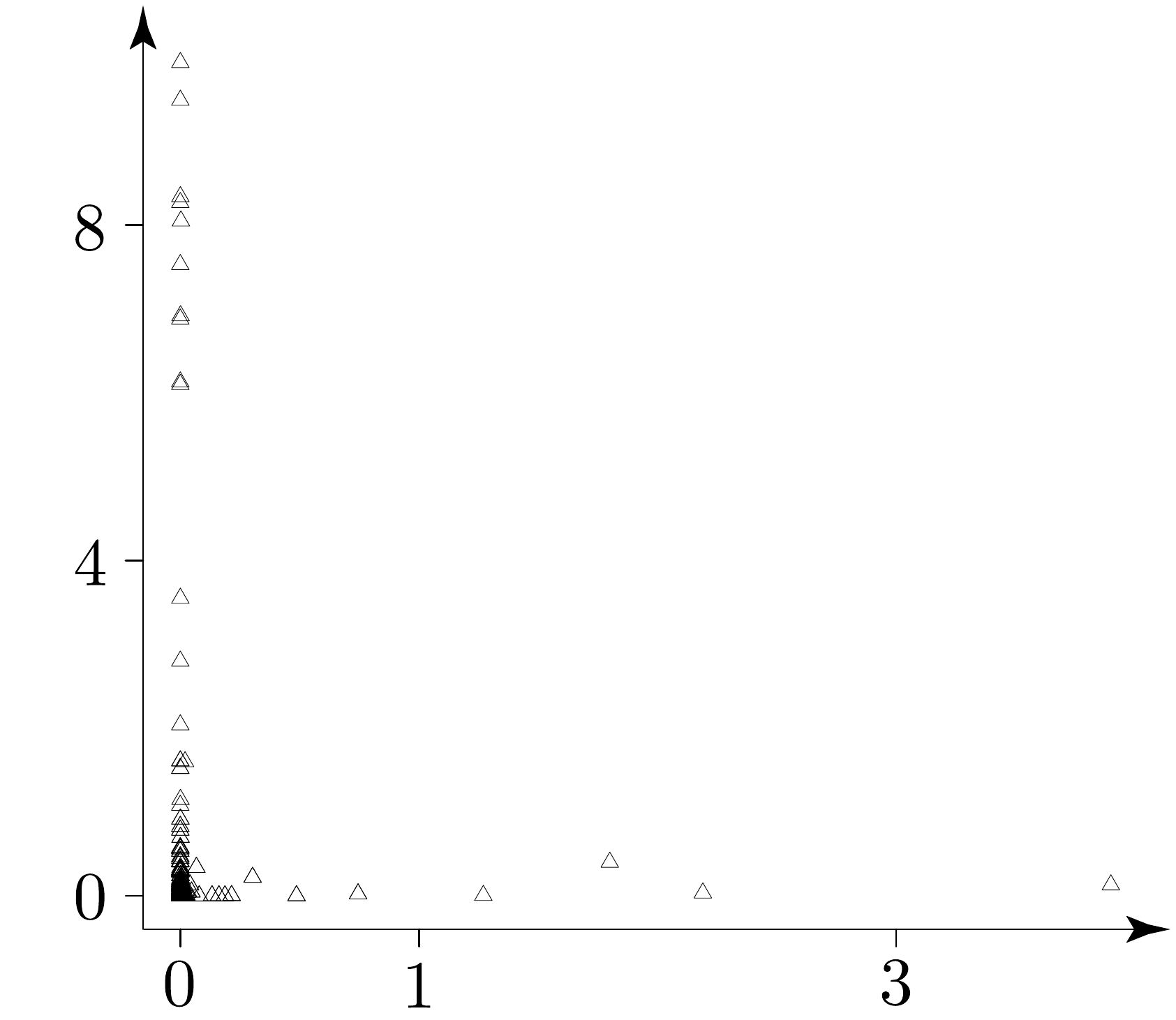}
\caption{}\label{fig:Pareto_60Hz_m5}
\end{subfigure}
\hspace*{\fill} 
\caption{Pareto fronts for $\omega=60~{\rm s}^{-1}$; plots~(a) to (e): first mode to fifth mode}
\label{fig:Pareto_60Hz_testII}
\end{figure}


\begin{figure} 
\hspace*{\fill} 
\begin{subfigure}{0.4\textwidth}
\includegraphics[width=\linewidth]{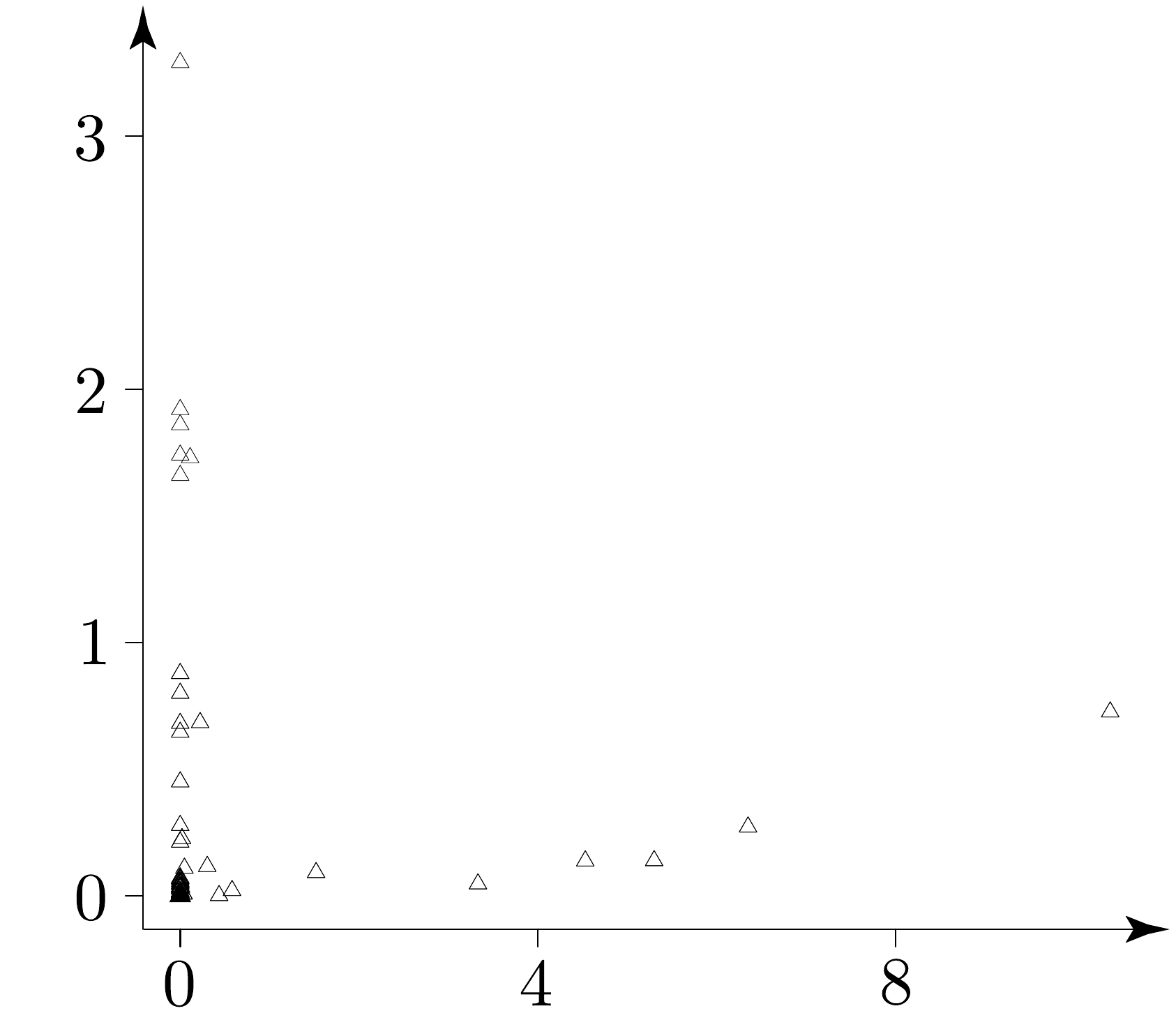}
\caption{}\label{fig:Pareto_100Hz_m1}
\end{subfigure}
\hspace*{\fill} 
\begin{subfigure}{0.4\textwidth}
\includegraphics[width=\linewidth]{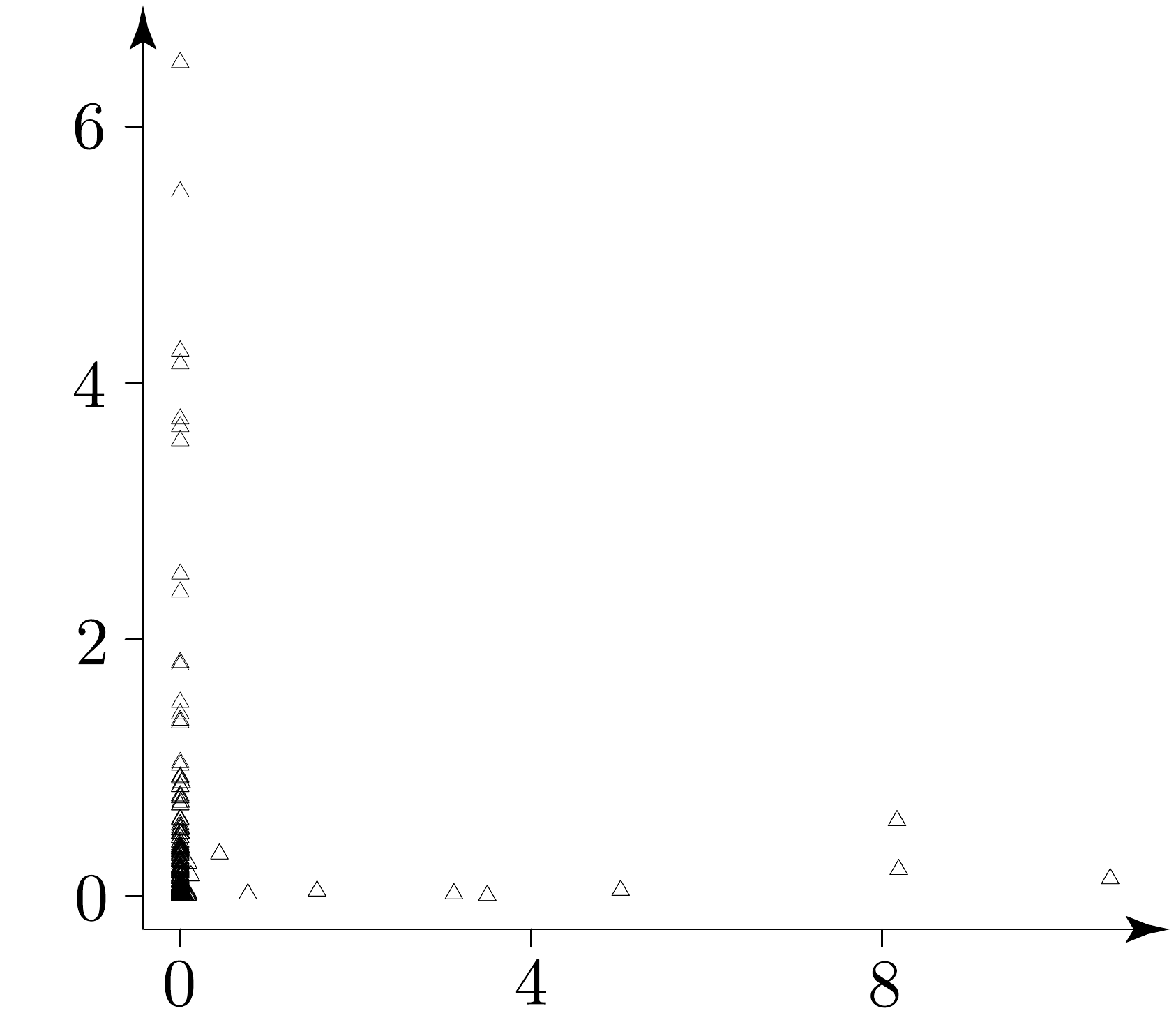}
\caption{}\label{fig:Pareto_100Hz_m2}
\end{subfigure}
\hspace*{\fill} 
\begin{subfigure}{0.4\textwidth}
\includegraphics[width=\linewidth]{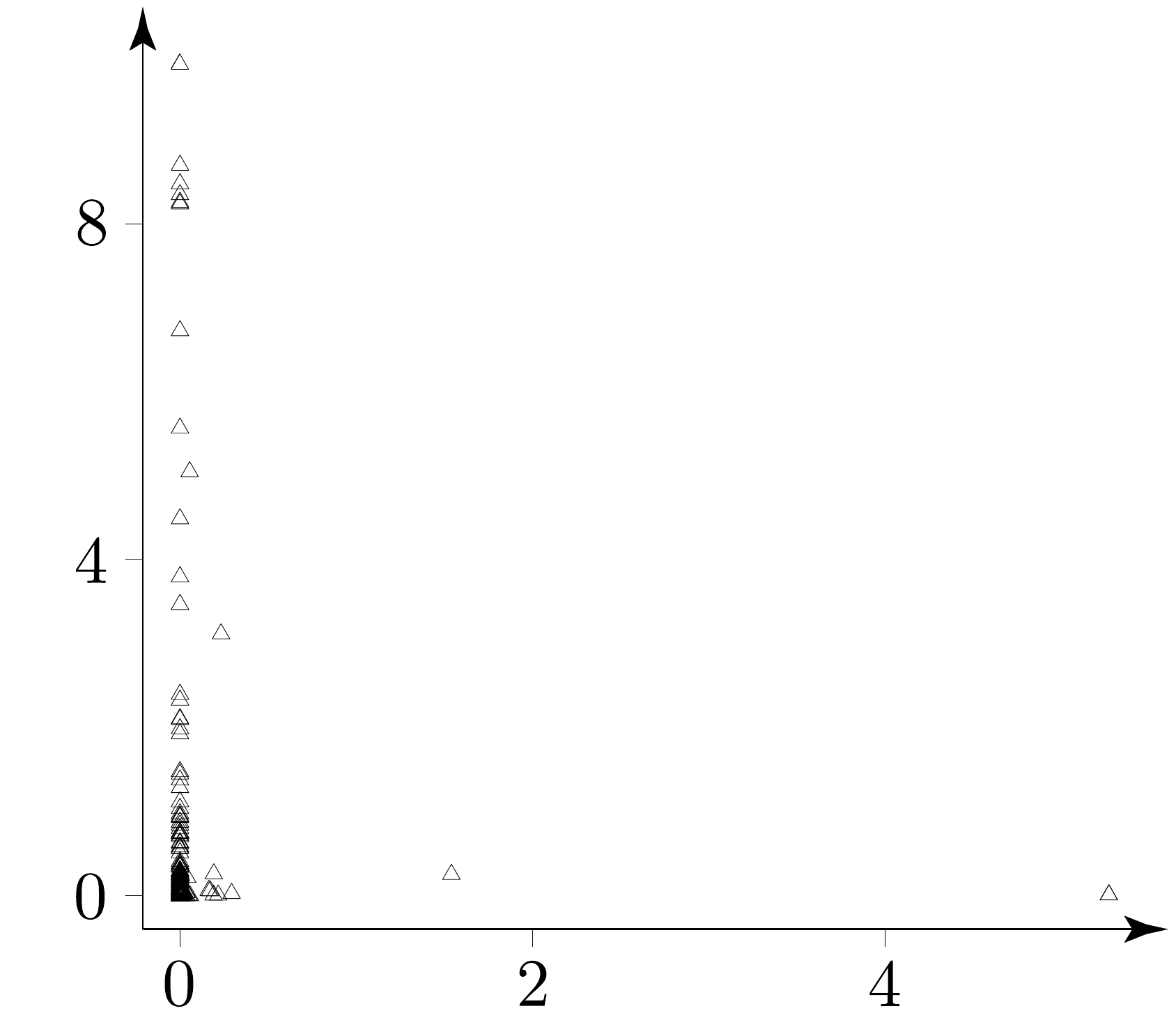}
\caption{}\label{fig:Pareto_100Hz_m3}
\end{subfigure}
\hspace*{\fill} 
\begin{subfigure}{0.4\textwidth}
\includegraphics[width=\linewidth]{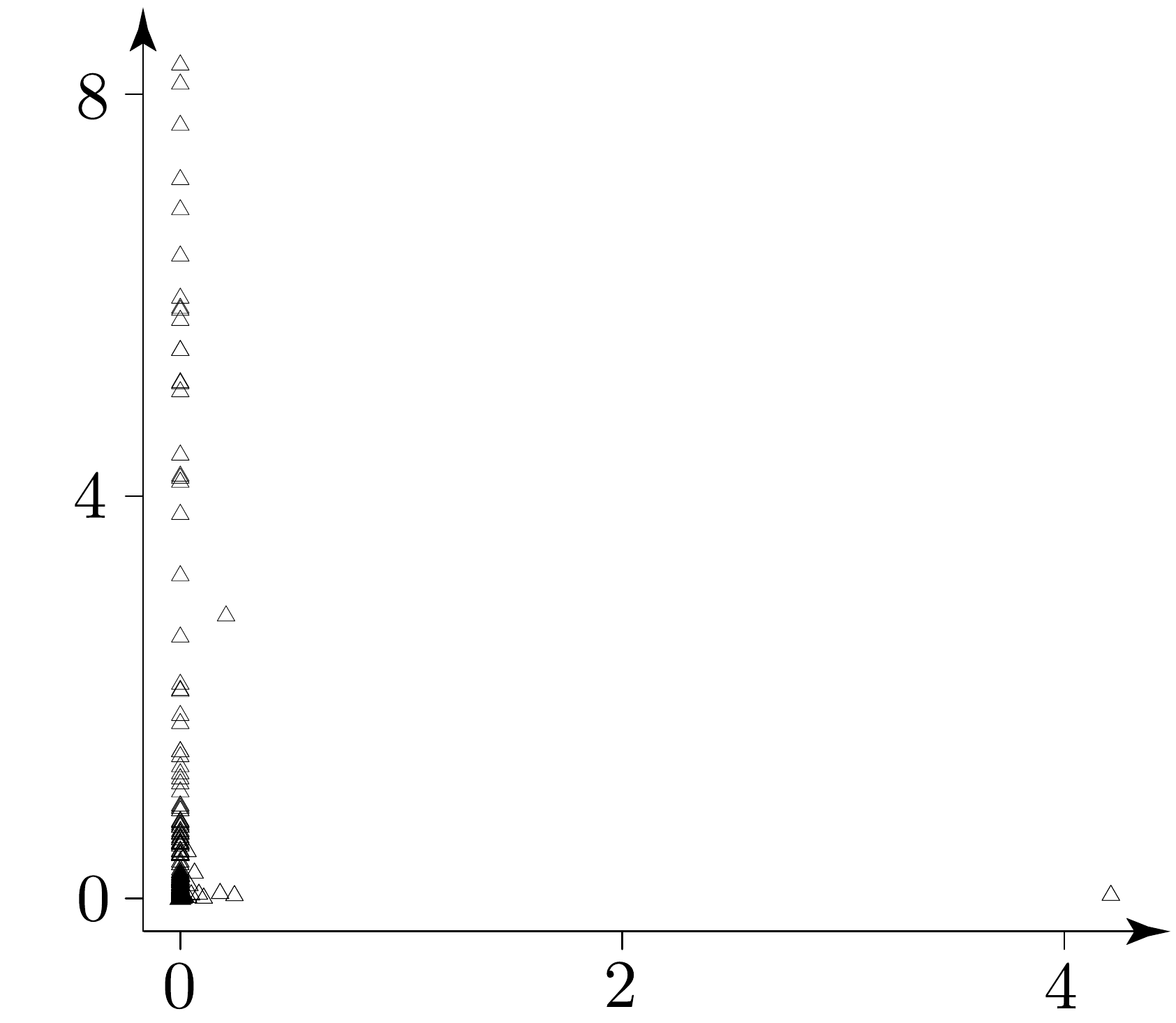}
\caption{}\label{fig:Pareto_100Hz_m4}
\end{subfigure}
\label{fig:Pareto_100Hz_a_testII}
\end{figure}
\addtocounter{figure}{-1}
\begin{figure} 
\hspace*{\fill} 
\begin{subfigure}{0.4\textwidth}
\addtocounter{subfigure}{4}
\includegraphics[width=\linewidth]{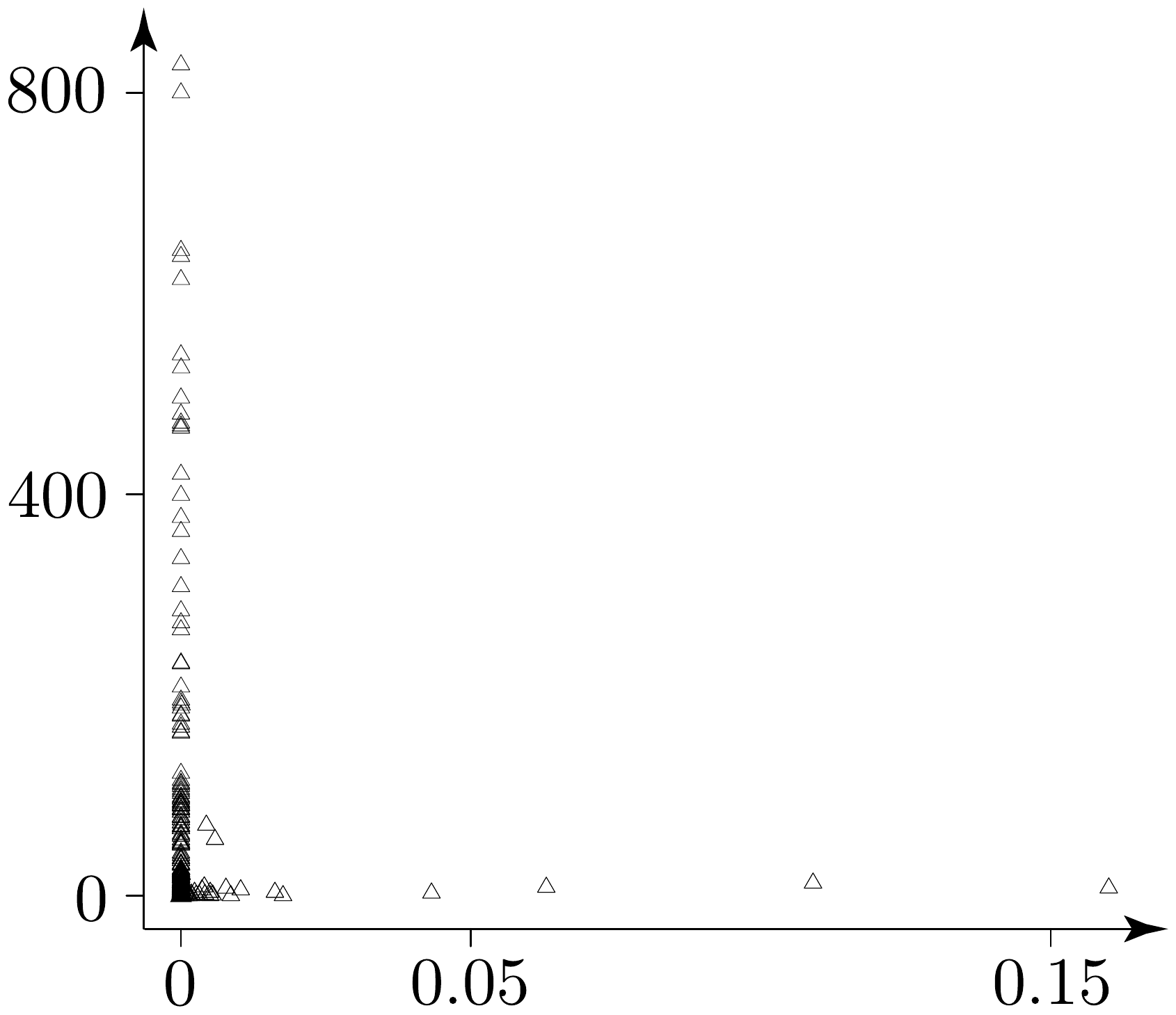}
\caption{}\label{fig:Pareto_100Hz_m5}
\end{subfigure}
\hspace*{\fill} 
\begin{subfigure}{0.4\textwidth}
\includegraphics[width=\linewidth]{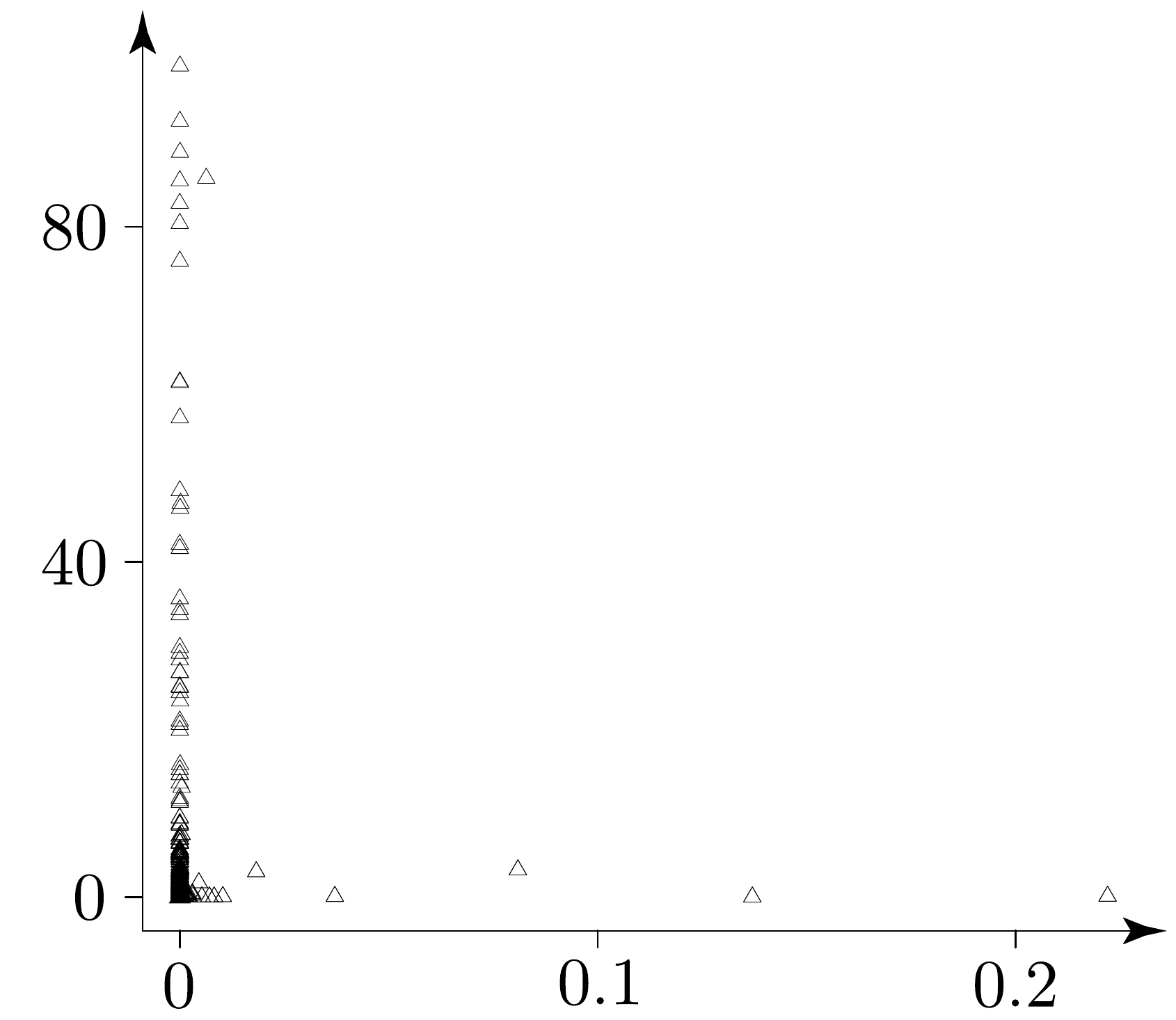}
\caption{}\label{fig:Pareto_100Hz_m6}
\end{subfigure}
\hspace*{\fill} 
\begin{subfigure}{0.4\textwidth}
\includegraphics[width=\linewidth]{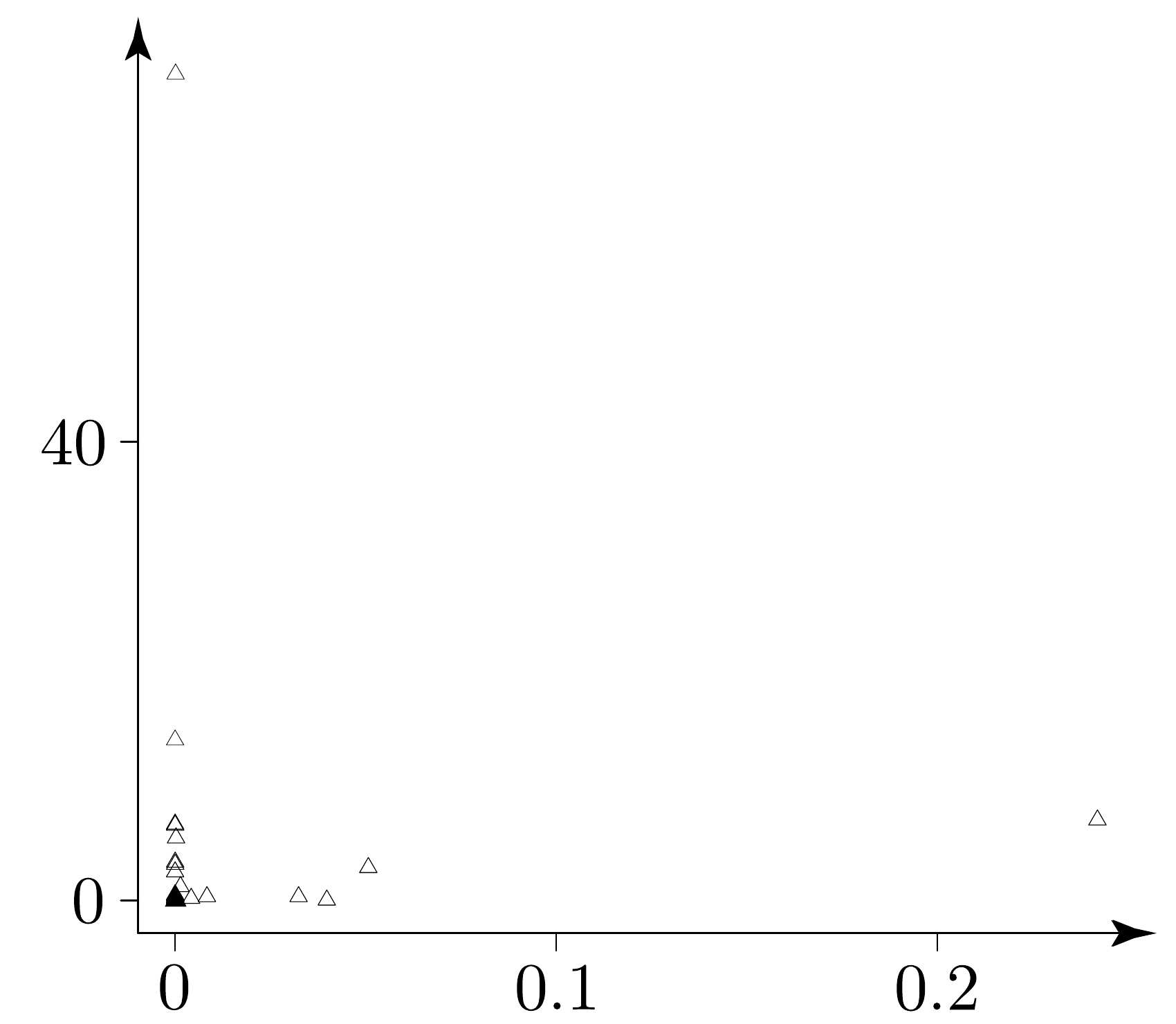}
\caption{}\label{fig:Pareto_100Hz_m7}
\end{subfigure}
\hspace*{\fill}
\\
\caption{Pareto fronts for $\omega=100~{\rm s}^{-1}$; plots~(a) to (g): first mode to seventh mode}
\label{fig:Pareto_100Hz_b_testII}
\end{figure}
Figures~\ref{fig:Pareto_15Hz_testII}--\ref{fig:Pareto_100Hz_b_testII} illustrate Pareto fronts.
The three figures are for $\omega=15~{\rm s}^{-1}$\,, $60~{\rm s}^{-1}$ and $100~{\rm s}^{-1}$\,, respectively.
The horizontal axes correspond to the misfit of the quasi-Rayleigh dispersion relation and the vertical axes to the misfit of the Love dispersion relation.

The values along these axes are to be scaled as follows.
Figure~\ref{fig:Pareto_15Hz_testII}:~$\times 10^{-5}$\,;
Figure~\ref{fig:Pareto_60Hz_testII} (a, b, c):~$\times 10^{-6}$\,;
Figure~\ref{fig:Pareto_60Hz_testII}  (d, e):~$\times10^{-5}$\,;
Figure~\ref{fig:Pareto_100Hz_b_testII}  (a): horizontal axis $\times 10^{-9}$\,, vertical axis $\times 10^{-5}$\,;
Figure~\ref{fig:Pareto_100Hz_b_testII}  (b): horizontal axis $\times 10^{-11}$\,, vertical axis $\times 10^{-5}$\,;
Figure~\ref{fig:Pareto_100Hz_b_testII}  (c): horizontal axis $\times 10^{-10}$\,, vertical axis$\times 10^{-5}$\,;
Figure~\ref{fig:Pareto_100Hz_b_testII}  (d): horizontal axis $\times 10^{-9}$\,, vertical axis $\times 10^{-5}$\,;
Figure~\ref{fig:Pareto_100Hz_b_testII}  (e):~$\times 10^{-7}$\,;
Figure~\ref{fig:Pareto_100Hz_b_testII}  (f, g):~$\times 10^{-6}$\,.
 
 In all cases, the fronts have rectangular shape and cover a relativity large range of values, which means that, while obtaining an optimal value for one target function is relatively easy, the model parameters producing this value can generate a wide range of values of the other function. 
 This emphasizes the need for a joint inversion to avoid the ambiguity resulting from inverting the dispersion relation for a single type of waves.
 In other words, the redundancy of information increases the trustworthiness of inferences.
 
Examining all frequencies and modes, we see that proposed solutions, marked by dark triangles, are more concentrated near~$(0,0)$\,, along the horizontal axis, which corresponds to quasi-Rayleigh waves, and more spread out along the vertical axis, which corresponds to the Love waves. 
It means that quasi-Rayleigh waves lend themselves particularly well to such an optimization and, if a satisfactory solution for Love waves is found, the quasi-Rayleigh target function can be easily adjusted.

Let us examine Table~\ref{tab:solutions_II}, which contains the actual and estimated values, as well as Table~\ref{tab:solutions_II_per}, where the ratio of the actual and estimated values is expressed in terms of percents.
The values of all parameters are inverted satisfactorily.
As in the case of a few outliers appearing in Figures~\ref{fig:Pareto_15Hz_testII}--\ref{fig:Pareto_100Hz_b_testII}, discrepancies might be caused by differences in positions of global minima of target functions and by occasional spurious results due to local minima and the numerical complexity the algorithm. 

\begin{table}
\caption{Summary of obtained results: Elasticity parameters are in units of $10^{10}{\rm N}/{\rm m}^2$\,, mass densities in $10^3{\rm kg}/{\rm m}^3$ and layer thickness in metres.}
\label{tab:solutions_II}
\begin{center}
{\small
\begin{tabular}{|c|c|c|c|c|c|c|c|c|c|}
  \hline     
         & $C_{11}^u$ & $C_{44}^u$ & $C_{11}^d$ & $C_{44}^d$ & $\rho^{u}$ & $\rho^{d}$ & $Z$\\
  \hline
Actual     & 1.980 & 0.880 & 10.985 & 4.160 & 2.200 & 2.600 & 500.0\\
  \hline
\multicolumn{8}{|c|}{$\omega=15~{\rm s}^{-1}$}\\
  \hline
2nd mode  & 2.211 & 0.875 & 10.958 & 4.215 & 2.234 & 2.605 & 522.7\\
  \hline
1st mode  & 1.999 &	0.893 & 10.405	& 3.919	& 2.254	& 2.578 & 480.3\\
  \hline
\multicolumn{8}{|c|}{$\omega=60~{\rm s}^{-1}$}\\
\hline
5th mode & 2.026	& 0.919	& 10.919	& 4.354	& 2.240	& 2.690 & 512.0\\
\hline
4th mode &	2.035	& 0.893	& 10.707	& 4.128	& 2.256	& 2.464 & 494.9\\
\hline
3rd mode &	2.022	& 0.878	& 11.307	& 4.258	& 2.184	& 2.743 & 505.1\\
\hline
2nd mode &	2.009	& 0.881	& 11.239	& 3.805	& 2.221	& 2.667 & 476.2\\
\hline
1st mode &	1.992	& 0.884	& 11.030	& 3.951	& 2.212	& 2.632 & 491.5\\
\hline
\multicolumn{8}{|c|}{$\omega=100~{\rm s}^{-1}$}\\
\hline
7th mode & 2.015	& 0.751	& 10.089	& 4.168	& 2.177	& 2.724 & 517.4\\
\hline
6th mode & 1.949	& 0.871	& 10.848	& 4.285	& 2.174	& 2.563 & 501.1\\
\hline
5th mode & 1.981	& 0.878	& 10.846	& 4.074	& 2.192	& 2.682 & 500.4\\
\hline
4th mode & 2.038	& 0.912	& 10.683	& 4.079	& 2.283	& 2.561 & 498.0\\
\hline
3rd mode & 1.954	& 0.872	& 10.882	& 4.210	& 2.174	& 2.538 & 506.1\\
\hline
2nd mode & 2.046	& 0.906	& 10.547	& 4.334	& 2.265	& 2.579 & 499.9\\
\hline
1st mode & 2.123	& 0.857	& 11.089	& 4.315	& 2.197	& 2.052 & 548.8\\
\hline
\end{tabular}}
\end{center}
\end{table}

\begin{table}
\caption{Estimated values compared to actual values, in percentages}
\label{tab:solutions_II_per}
\begin{center}
\begin{tabular}{|c|c|c|c|c|c|c|c|c|c|}
  \hline     
         & $C_{11}^u$ & $C_{44}^u$ & $C_{11}^d$ & $C_{44}^d$ & $\rho^{u}$ & $\rho^{d}$ & $Z$\\
  \hline		
   
Actual     & 100.0  & 100.0  & 100.0  & 100.0  & 100.0  & 100.0  & 100.0 \\
  \hline																																										
															
\multicolumn{8}{|c|}{$\omega=15~{\rm s}^{-1}$}\\																
\hline																
2nd	mode	& 111.7  & 99.4  & 99.8  & 101.3  & 101.6  & 100.2  & 104.5  \\
\hline																
1st	mode	& 100.9  & 101.5  & 94.7  & 94.2  & 102.5  & 99.2  & 96.1  \\
\hline																
\multicolumn{8}{|c|}{$\omega=60~{\rm s}^{-1}$}\\																
\hline																
5th	mode	& 102.3  & 104.4  & 99.4  & 104.7  & 101.8  & 103.4  & 102.4  \\
\hline																
4th	mode	& 102.8  & 101.5  & 97.5  & 99.2  & 102.6  & 94.8  & 99.0  \\
\hline																
3rd	mode	& 102.1  & 99.8  & 102.9  & 102.4  & 99.3  & 105.5  & 101.0  \\
\hline																
2nd	mode	& 101.5  & 100.1  & 102.3  & 91.5  & 101.0  & 102.6  & 95.2  \\
\hline																
1st	mode	& 100.6  & 100.5  & 100.4  & 95.0  & 100.5  & 101.2  & 98.3  \\
\hline																
\multicolumn{8}{|c|}{$\omega=100~{\rm s}^{-1}$}\\																
\hline																
7th	mode	& 101.8  & 85.3  & 91.8  & 100.2  & 99.0  & 104.8  & 103.5  \\
\hline																
6th	mode	& 98.4  & 99.0  & 98.8  & 103.0  & 98.8  & 98.6  & 100.2  \\
\hline																
5th	mode	& 100.0  & 99.7  & 98.7  & 97.9  & 99.6  & 103.1  & 100.1  \\
\hline																
4th	mode	& 102.9  & 103.6  & 97.3  & 98.1  & 103.8  & 98.5 & 99.6   \\
\hline																
3rd	mode	 & 98.7  & 99.1  & 99.1  & 101.2  & 98.8  & 97.6  & 101.2 \\
\hline																
2nd	mode	& 103.3  & 103.0  & 96.0  & 104.2  & 103.0  & 99.2  & 100.0  \\
\hline																
1st	mode	& 107.2  & 97.4  & 100.9  & 103.7  & 99.8  & 78.9  & 109.8  \\
\hline	
\end{tabular}
\end{center}
\end{table}

\begin{figure}
\includegraphics[width=\linewidth]{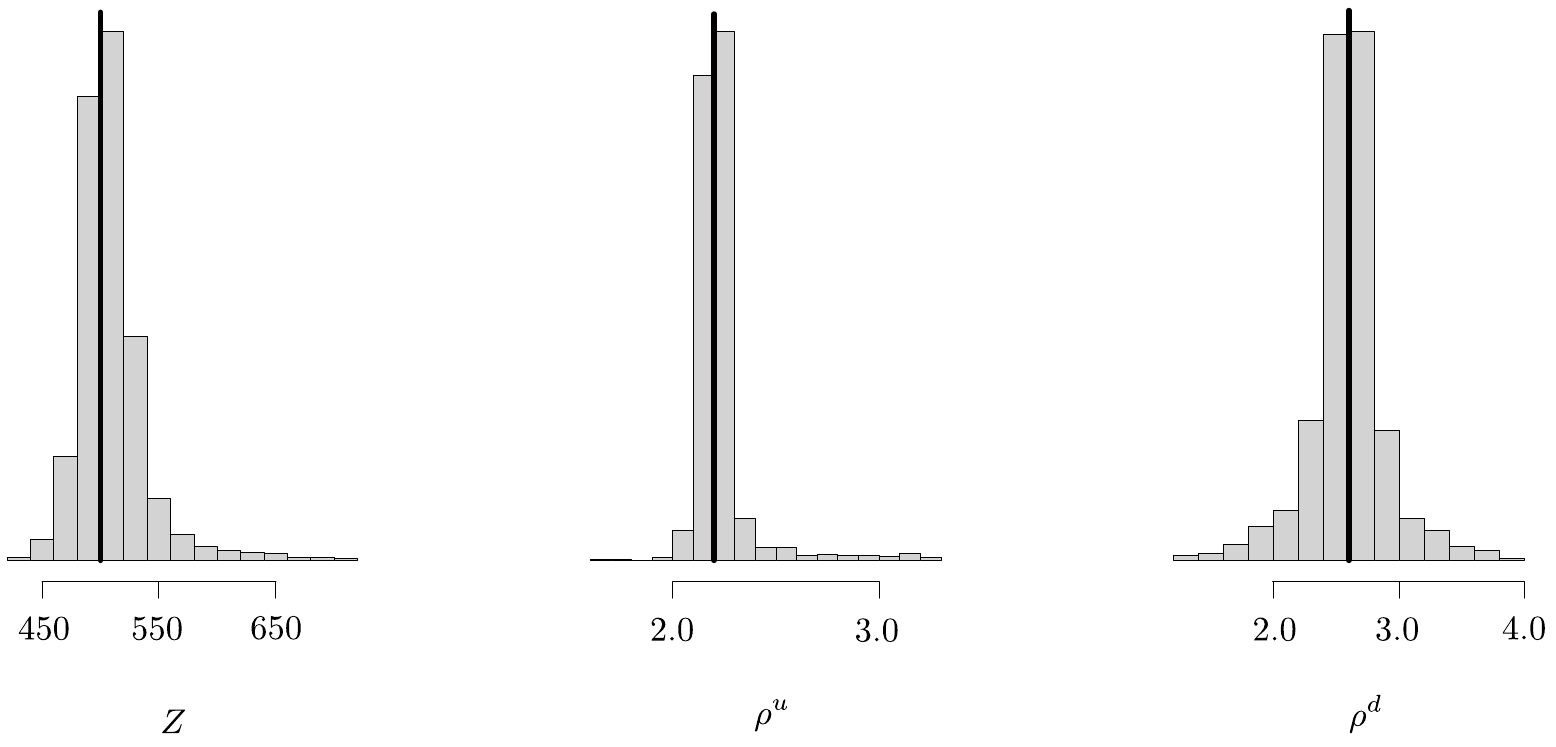}
\caption{Layer thickness, in metres, and the layer and halfspace mass densities, in $10^3$ ${\rm kg}/{\rm m}^3$\,; black lines represent the actual values}
\label{fig:h_1}
\end{figure}

\begin{figure}
\includegraphics[width=\linewidth]{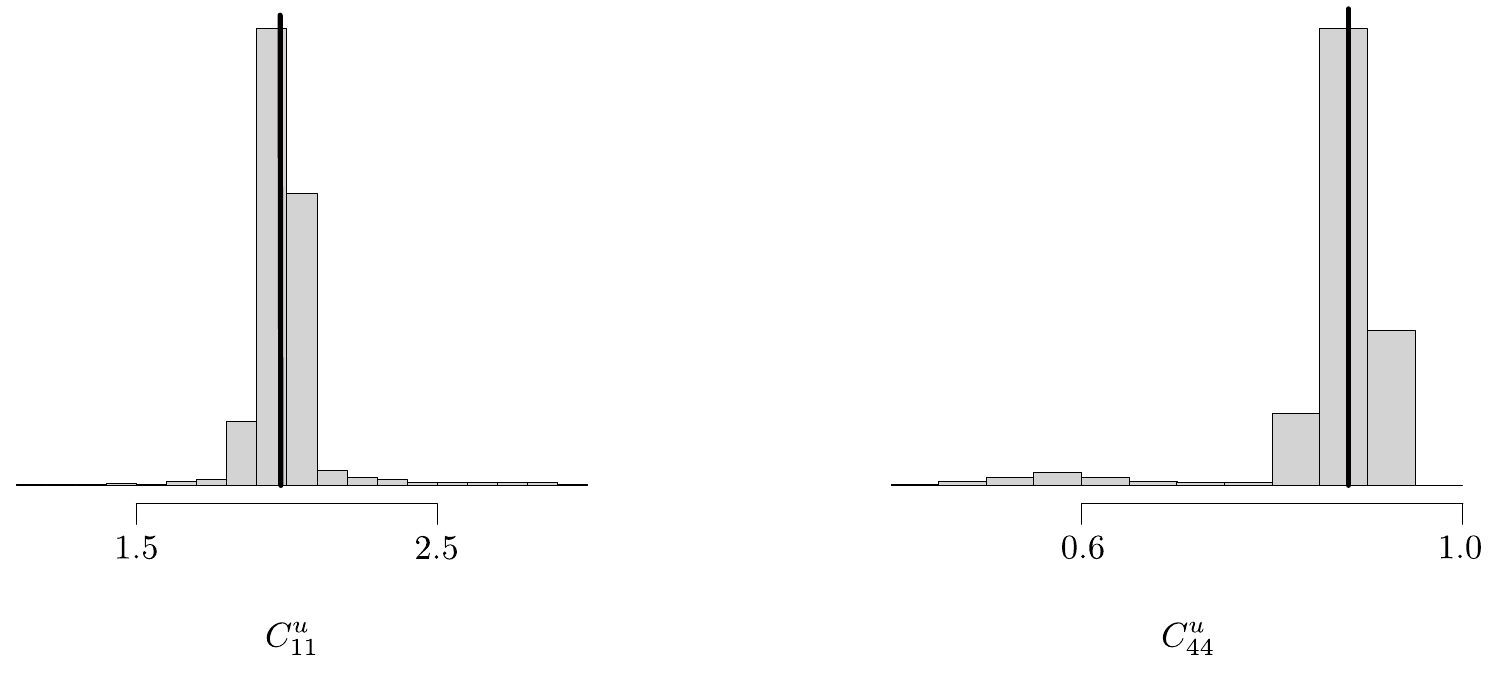}
\caption{Layer elasticity parameters, in $10^{10}$ ${\rm N}/{\rm m}^2$\,; black lines represent actual values}
\label{fig:h_2}
\end{figure}

\begin{figure}
\includegraphics[width=\linewidth]{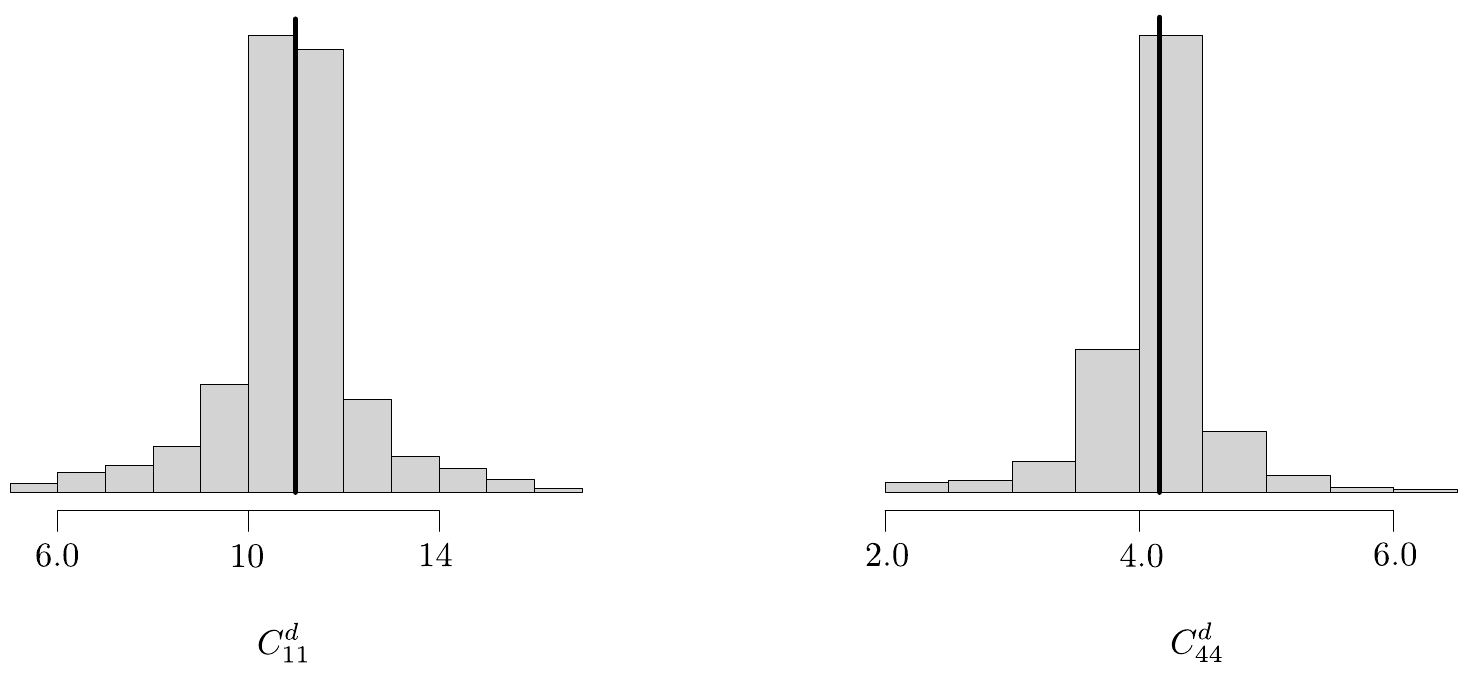}
\caption{Halfspace elasticity parameters, in $10^{10}$ ${\rm N}/{\rm m}^2$\,; black lines represent actual values}
\label{fig:h_3}
\end{figure}

Figures~\ref{fig:h_1}--\ref{fig:h_3} are histograms of model parameters obtained along the Pareto fronts for the fundamental mode at $\omega=60~{\rm s}^{-1}$\,.   
The spread in values is not due to perturbations, as is commonly the case for histograms, but instead shows the range of Pareto optimal solutions along the Pareto front.
For each parameter there is a good match between the values obtained by the inverse process and the values used in the original dispersion relations.
Also, the aforementioned spread is narrow, as expected in view of the concentration of solutions near~$(0,0)$\,, in Figures~\ref{fig:Pareto_15Hz_testII}--\ref{fig:Pareto_100Hz_b_testII}.
For higher modes and different frequencies, the results are similar to the ones illustrated in Figures~\ref{fig:h_1}--\ref{fig:h_3}, as can be inferred from Tables~\ref{tab:solutions_II} and \ref{tab:solutions_II_per}.

\begin{table}
\caption{Summary of obtained results for input errors of $\pm 5\%$\,.  Elasticity parameters are in units of $10^{10}{\rm N}/{\rm m}^2$\,, mass densities in $10^3{\rm kg}/{\rm m}^3$ and layer thickness in metres.}
\label{tab:solutions_+5per}
\begin{center}
\begin{tabular}{|c|c|c|c|c|c|c|c|c|c|}
  \hline     
         & $C_{11}^u$ & $C_{44}^u$ & $C_{11}^d$ & $C_{44}^d$ & $\rho^{u}$ & $\rho^{d}$ & $Z$\\
  \hline
Actual    & 1.980 & 0.880 & 10.985 & 4.160 & 2.200 & 2.600 & 500.0\\
  \hline
\multicolumn{8}{|c|}{$\omega=15~{\rm s}^{-1}$}\\
  \hline
2nd mode  & 1.899 & 0.856 & 10.440 & 4.002 & 2.296 & 2.748 & 509.4\\
  \hline
1st mode  & 2.121 & 0.943 & 10.736 & 4.061 & 2.135 & 2.435 & 524.4\\
  \hline
\multicolumn{8}{|c|}{$\omega=60~{\rm s}^{-1}$}\\
\hline
5th mode & 2.312 & 0.899 & 11.017 & 3.861 & 2.091 & 2.679 & 535.7\\
\hline
4th mode & 2.178 & 0.859 & 9.541 & 3.969 & 2.022 & 2.593 & 504.1\\
\hline
3rd mode & 1.921 & 0.917 &	11.139 & 3.992 & 2.119 & 2.231 & 505.9\\
\hline
2nd mode & 1.711 & 0.843 & 10.956 & 4.328 & 2.296 & 2.522 & 523.5\\
\hline
1st mode & 2.113 &	0.933 & 15.211 &	4.004 & 2.120 &	2.929 & 482.9\\ 
\hline
\multicolumn{8}{|c|}{$\omega=100~{\rm s}^{-1}$}\\
\hline
7th mode & 2.219 & 0.890 & 10.854 & 3.753 & 1.263 & 2.660 & 527.2\\
\hline
6th mode & 1.907 & 0.863 & 9.640 & 4.177 & 2.177 & 2.024 & 530.0\\
\hline
5th mode & 2.495 & 0.849 & 11.556 & 4.208 & 2.246 & 2.492 & 513.9\\
\hline
4th mode & 1.952 & 0.923 &	12.543 & 4.031 & 2.117 & 2.665 & 508.5\\
\hline
3rd mode & 1.937 & 0.864 & 10.480 & 4.291 & 2.239 & 2.549 & 491.5\\
\hline
2nd mode & 2.066 & 0.837 & 11.757 & 4.232 & 2.322 & 2.463 & 470.0\\
\hline
1st mode & 2.068 & 0.863 & 11.633 & 4.353 & 2.488 & 2.613 & 560.9\\
\hline
\end{tabular}
\end{center}
\end{table}

\begin{table}
\caption{Estimated values compared to actual values, in percentages, for input errors of $\pm 5\%$}
\label{tab:solutions_per_5}
\begin{center}
\begin{tabular}{|c|c|c|c|c|c|c|c|c|c|}
  \hline     
          & $C_{11}^u$ & $C_{44}^u$ & $C_{11}^d$ & $C_{44}^d$ & $\rho^{u}$ & $\rho^{d}$ & $Z$\\
  \hline		
   
Actual     & 100.0  & 100.0  & 100.0  & 100.0  & 100.0  & 100.0  & 100.0 \\
  \hline																																										
															
\multicolumn{8}{|c|}{$\omega=15~{\rm s}^{-1}$}\\																					
\hline																
2nd	mode	& 95.9	& 97.2	& 95.08	& 96.28	& 104.4	& 105.7	& 101.9\\
\hline																
1st	mode	& 107.1& 107.2	& 97.7	& 97.6	& 97.0	& 93.7	& 104.9\\
\hline																
\multicolumn{8}{|c|}{$\omega=60~{\rm s}^{-1}$}\\																				
\hline																
5th	mode	& 116.8	& 102.1	& 100.3	& 92.8	& 95.0	& 103.0	& 107.1\\
\hline																
4th	mode	& 110.0	& 97.6	& 86.9	& 95.4	& 91.9	& 99.7	& 100.8\\
\hline																
3rd	mode	& 97.0	& 104.2	& 101.4	& 96.0	& 96.3	& 85.8	& 101.2\\
\hline																
2nd	mode	& 86.4	& 95.8	& 99.7	& 104.0	& 104.4	& 97.0	& 104.7\\
\hline																
1st	mode	& 106.7	& 106.1 & 138.5	& 96.2	& 96.4	& 112.6	& 96.6\\
\hline																
\multicolumn{8}{|c|}{$\omega=100~{\rm s}^{-1}$}\\																	
\hline																
7th	mode	& 112.1	& 101.2& 98.8	& 90.2	& 57.4	& 102.3	& 105.4\\
\hline																
6th	mode	& 96.3	& 98.0	& 87.8	& 100.4	& 97.0	& 77.8	& 106.0\\
\hline																
5th	mode	& 126.0	& 96.4	& 105.2	& 101.1	& 102.1& 95.9	& 102.8\\
\hline																
4th	mode	& 98.6	& 104.9	& 114.2	& 96.9	& 96.2	& 102.5	& 101.7\\
\hline																
3rd	mode	& 97.8	& 98.1	& 95.4	& 103.1	& 101.8	& 98.0	& 98.3\\
\hline																
2nd	mode	& 104.4	& 95.2	& 107.0	& 101.7	& 105.5	& 94.7	& 94.0\\
\hline																
1st	mode	& 104.5& 98.0	& 105.9	& 104.6	& 113.1	& 100.5	& 112.2\\
\hline	
\end{tabular}
\end{center}
\end{table}

\begin{figure}
\centerline{
\includegraphics[scale=0.47]{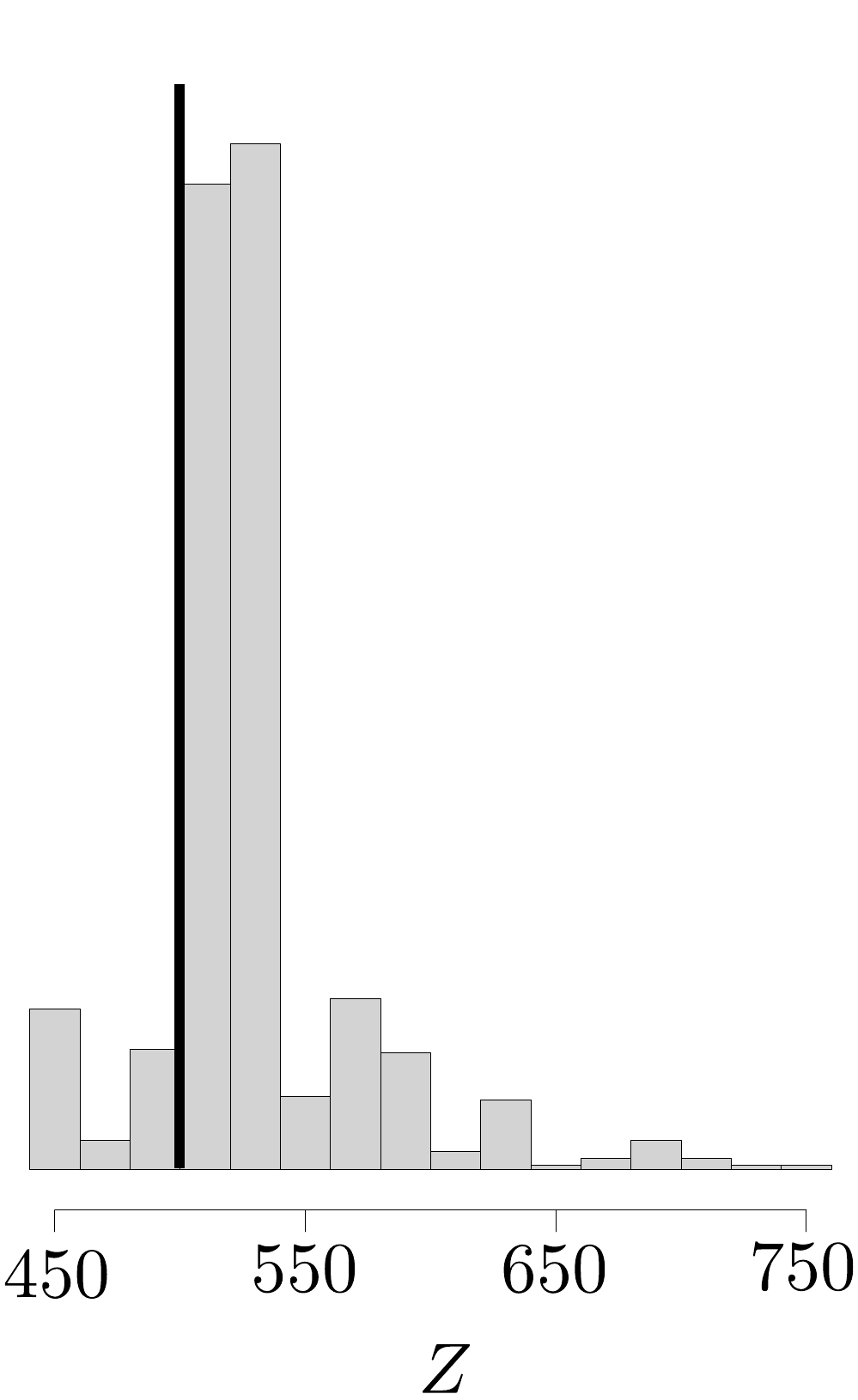}\hspace{0.5in}
\includegraphics[scale=0.47,trim=0 4 0 0]{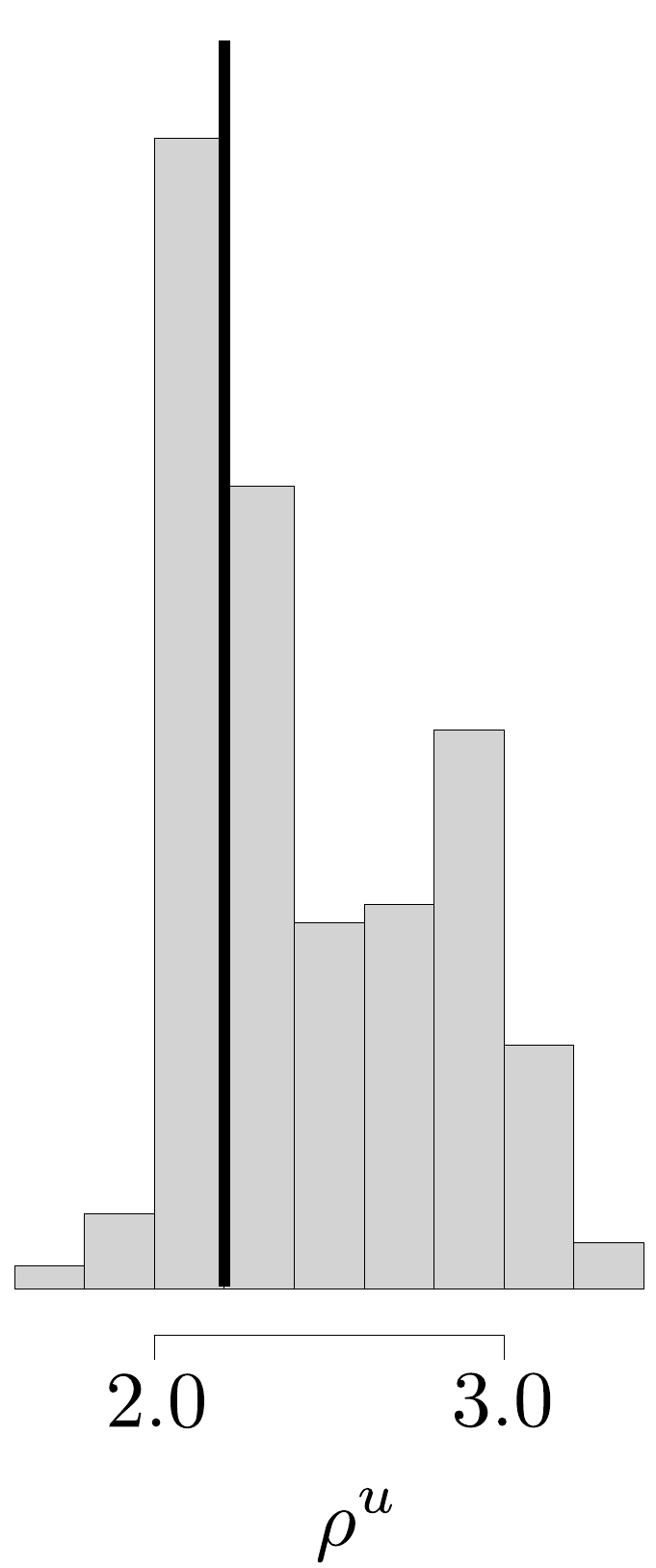}\hspace{0.5in}
\includegraphics[scale=0.47,trim=0 4 0 0]{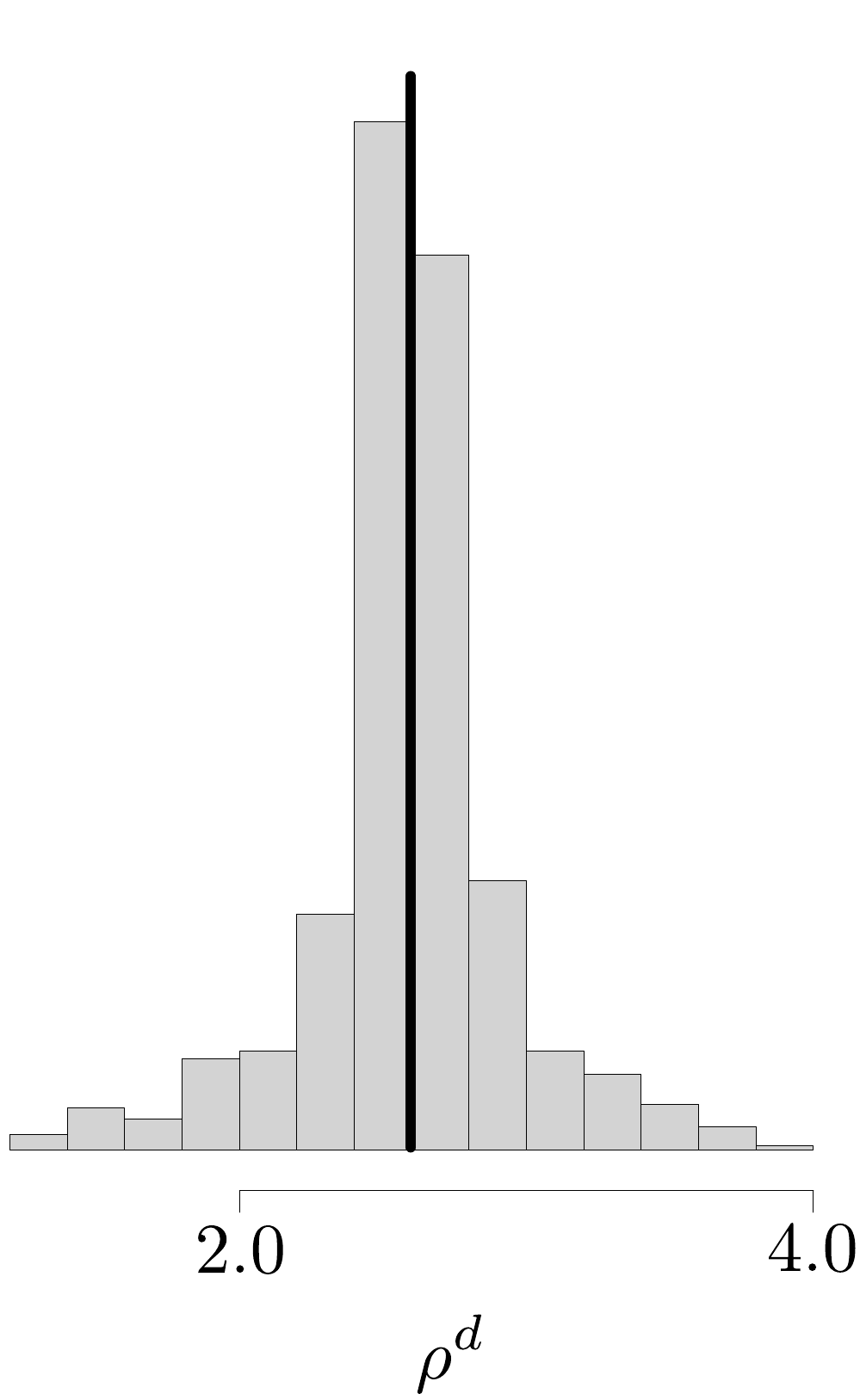}
}
\caption{Layer thickness, in metres, and the layer and halfspace mass densities, in $10^3$ ${\rm kg}/{\rm m}^3$\,, for the input error of $+5\%$\,; black lines represent the actual values}
\label{fig:h_4}
\end{figure}

\begin{figure}
\centerline{
\includegraphics[scale=0.47]{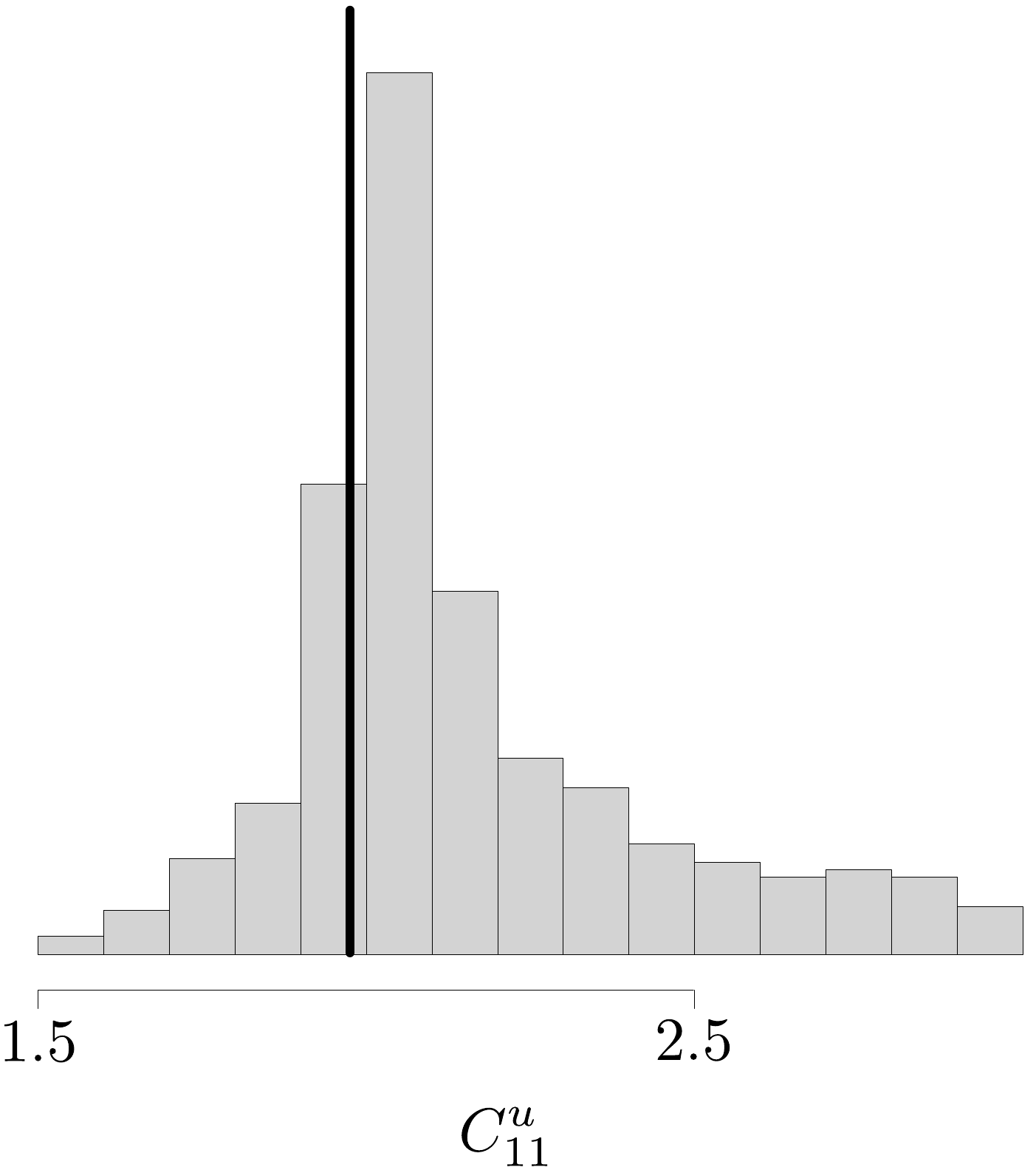}\hspace{0.6in}
\includegraphics[scale=0.47]{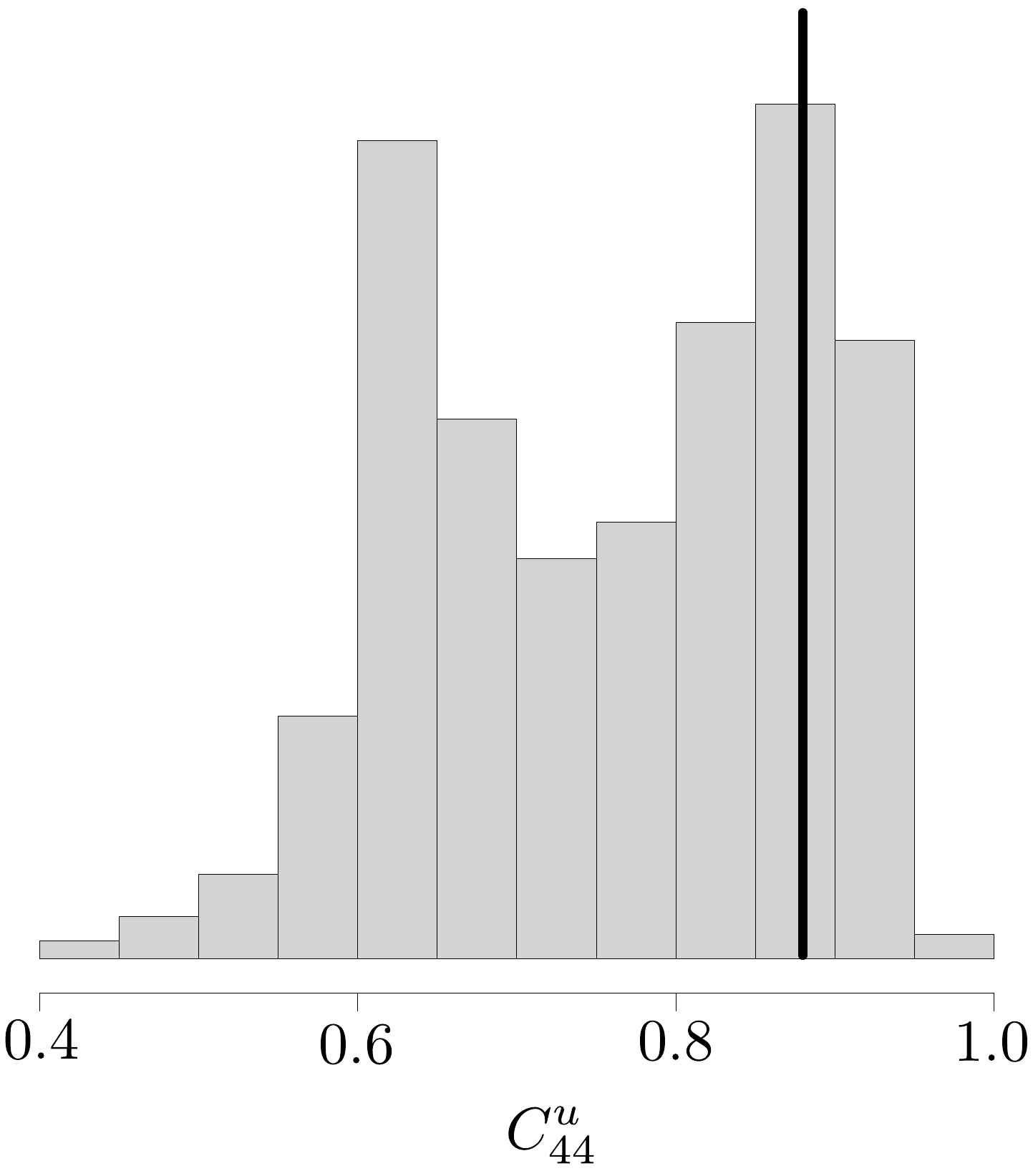}
}\caption{Layer elasticity parameters, in $10^{10}$ ${\rm N}/{\rm m}^2$\,, for the input error of $+5\%$; black lines represent actual values}
\label{fig:h_5}
\end{figure}

\begin{figure}
\centerline{
\includegraphics[scale=0.47]{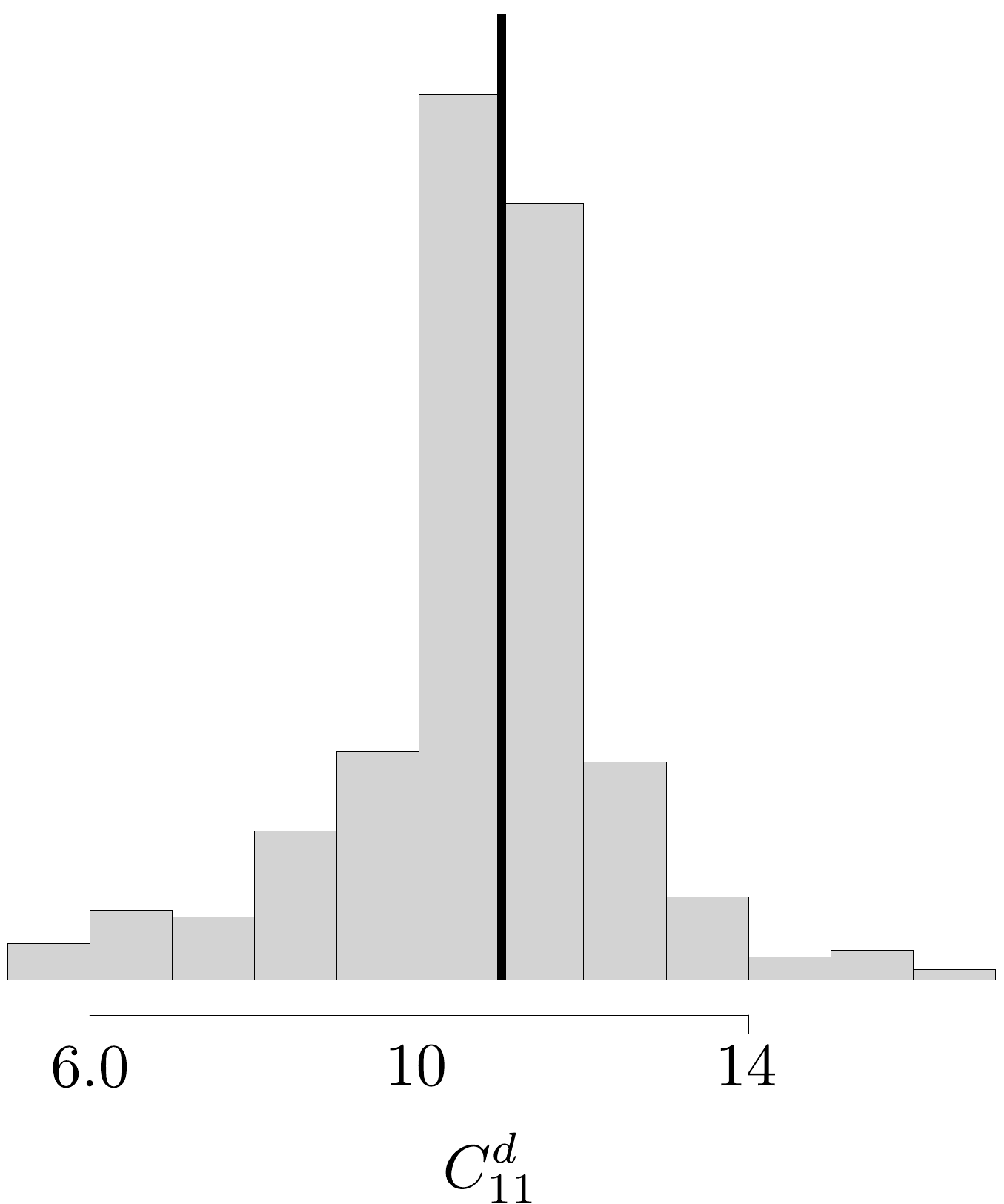}\hspace{0.6in}
\includegraphics[scale=0.47]{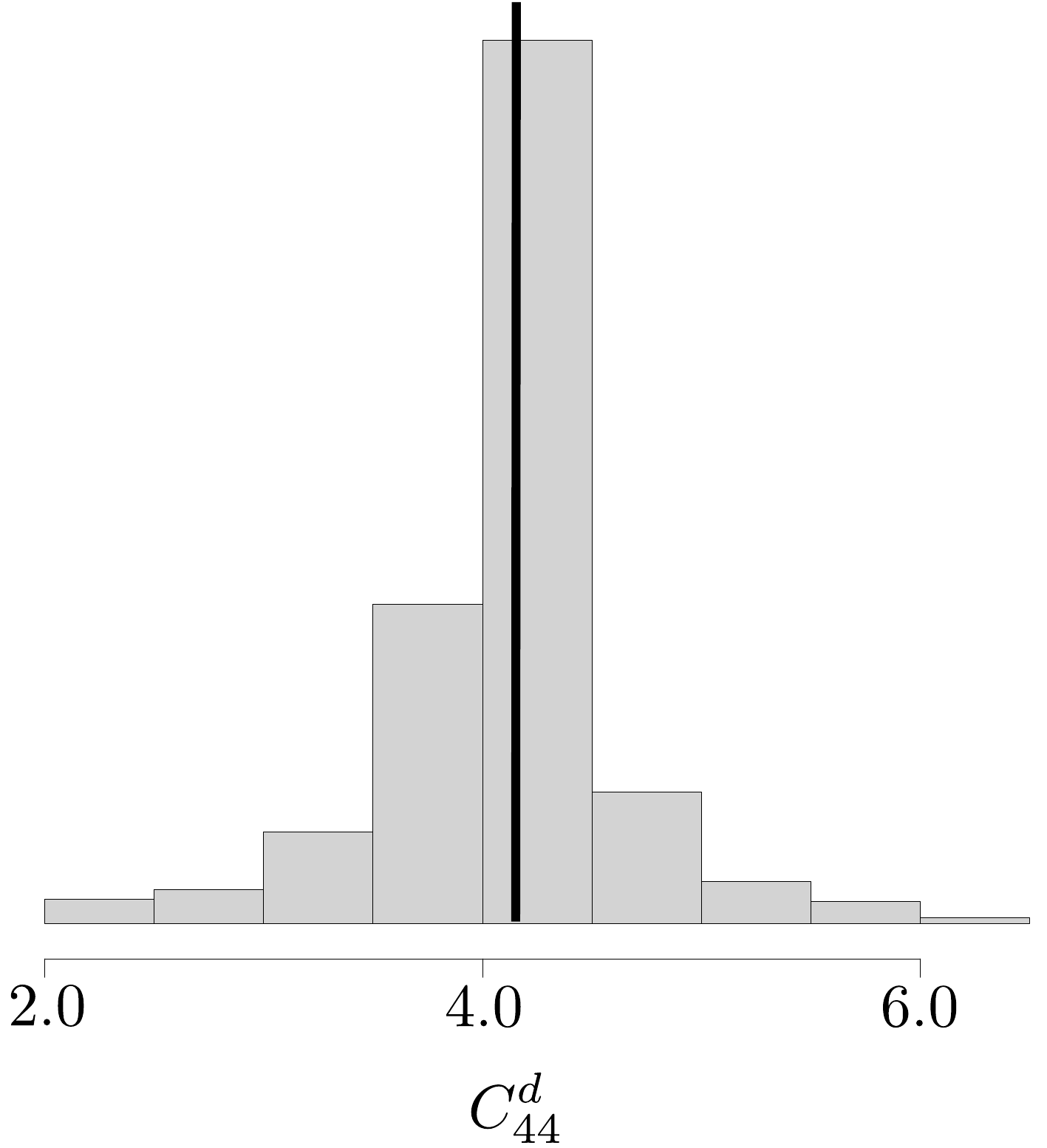}
}\caption{Halfspace elasticity parameters, in $10^{10}$ ${\rm N}/{\rm m}^2$\,, for the input error of $+5\%$\,; black lines represent actual values}
\label{fig:h_6}
\end{figure}
\subsection{Input with errors}
We wish to perform an uncertainty analysis of the Pareto joint inversion. 
However, in the case of guided waves, standard perturbation methods are not feasible. 
First of all, a single front is already a collection of hundreds of individual solutions, so a perturbation approach would lead to parameter distributions expressing both the Pareto-front and perturbation effects. 
An interpretation of such a combination might be exceedingly difficult. 
The other problem is pragmatic; obtaining perturbed results for all frequencies and modes would require enormous computational time, perhaps even years. 

Thus, to examine the sensitivity of the proposed method, for all modes and frequencies, we examine the effects of $5\%$ errors in propagation speeds of the quasi-Rayleigh and Love waves. 
This allows us to gain an insight into effects of inaccuracies of the input on the estimation of model parameter, without performing a complete perturbation study.

Tables~\ref{tab:solutions_+5per} and~\ref{tab:solutions_per_5} contain the best solutions from the combination of all Pareto-optimal solutions for input errors of $+5\%$ and $-5\%$\,.
These values are to be compared with Tables~\ref{tab:solutions_II} and~\ref{tab:solutions_II_per}, which contain the best solutions obtained from the error-free input.
Figures~\ref{fig:h_4}--\ref{fig:h_6} are histograms of model parameters along the Pareto fronts for the fundamental mode at $\omega=60~{\rm s}^{-1}$\,, obtained with input error of $+5\%$\,.
These histograms are to be compared with Figures~\ref{fig:h_1}--\ref{fig:h_3}\,, which are obtained from error-free input.

Examining the tables, we see that even $5\%$ errors lead to a significant loss of accuracy.
Examining the histograms, we see that the peak values are shifted from the actual values, and the histograms have a greater spread, especially for $\rho^u$ and $C_{44}^u$\,.
To explain the last statement, consider the fact that---as seen in Figure~\ref{fig:dispersion} and as discussed by Ud\'{\i}as~\cite{udias}---the fundamental-mode dispersion curve of the quasi-Rayleigh wave is asymptotic to the propagation speed of a Rayleigh wave in a halfspace with the same properties as the layer, which is affected by the shear-wave speed in the layer,~$\beta^u=\sqrt{C_{44}^u/\rho^u}$\,.
Thus, an error in $v_r$ affects particularly the inverted values of $\rho^u$ and $C_{44}^u$\,.

We also note that, while the peak values of the histograms are those for which there are the most Pareto-optimal solutions, those values are not necessarily the best solutions.
For example, the histogram for $C_{11}^d$\,, in Figure~\ref{fig:h_6}\,, has a peak value around $11$\,, but the corresponding best value in Table~\ref{tab:solutions_+5per} is $15.211$\,.
\subsection{Correlations between elasticity parameters}
\begin{figure}
\centerline{
\hspace*{-0.1in}
\includegraphics[scale=0.42]{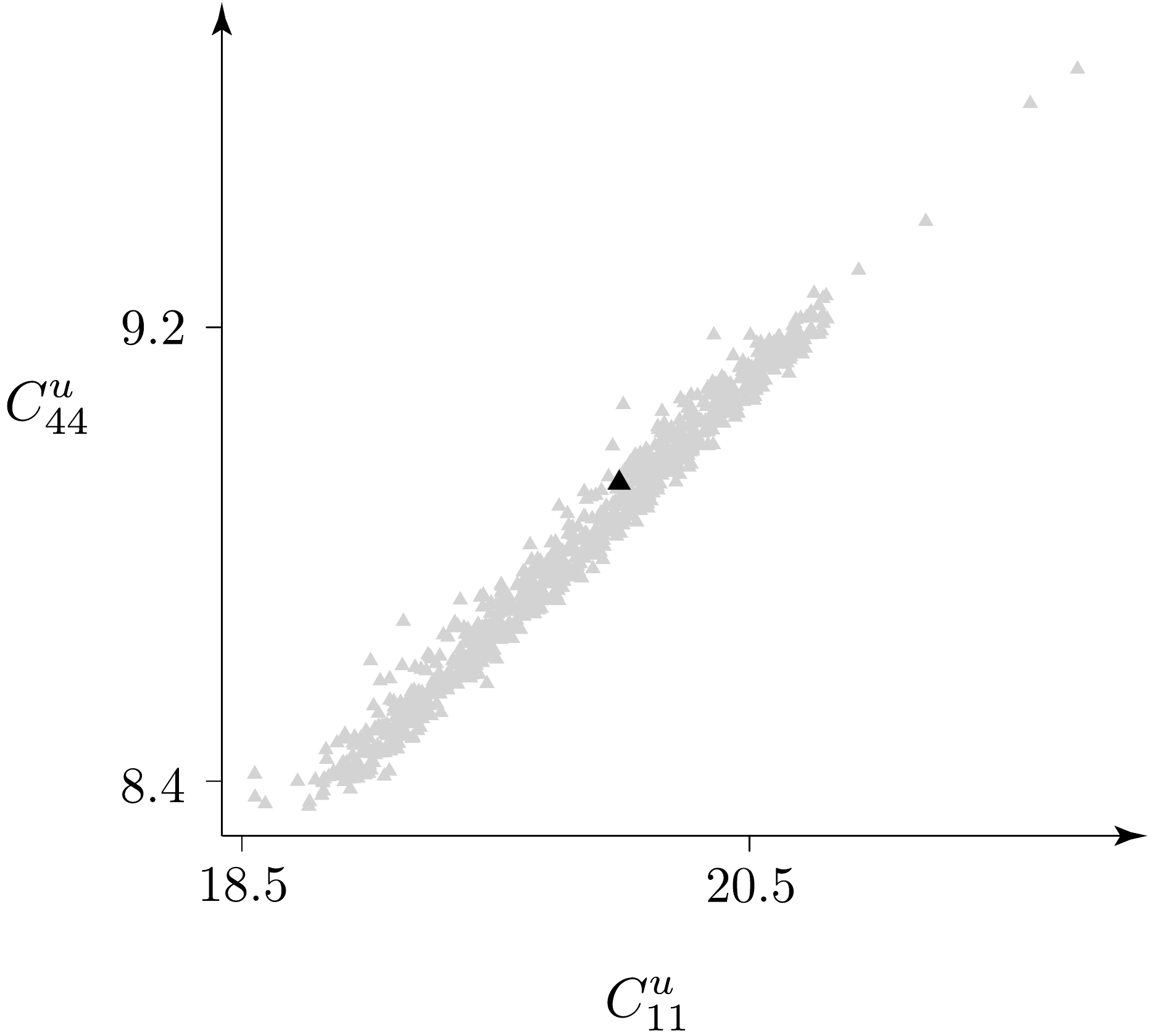}\hspace{0.3in}
\includegraphics[scale=0.42]{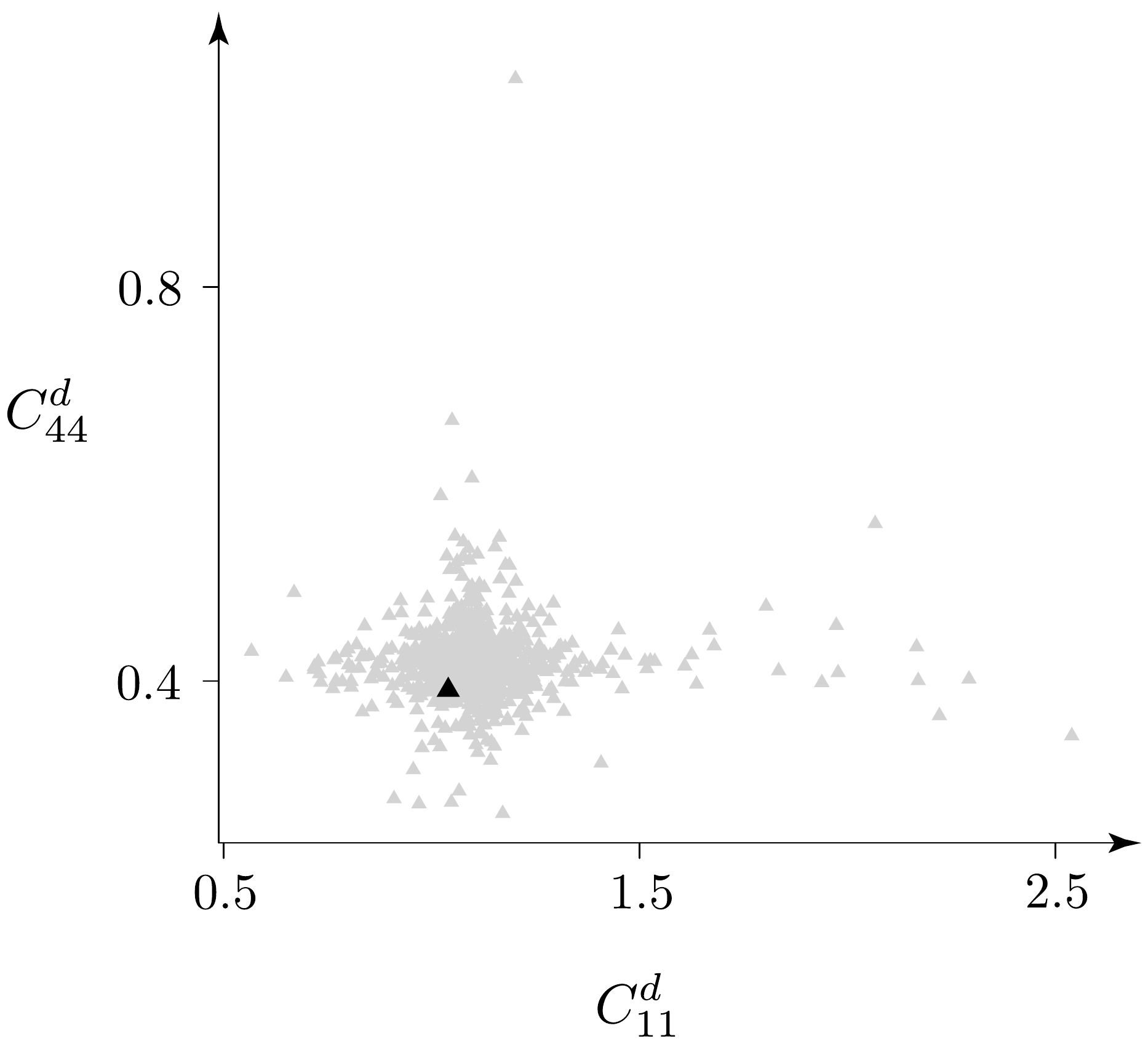}
}
\caption{Relationships between elasticity parameters for input without errors}
\label{fig:corr_unperturbed}
\end{figure}
\begin{figure}
\centerline{
\hspace*{-0.1in}
\includegraphics[scale=0.42]{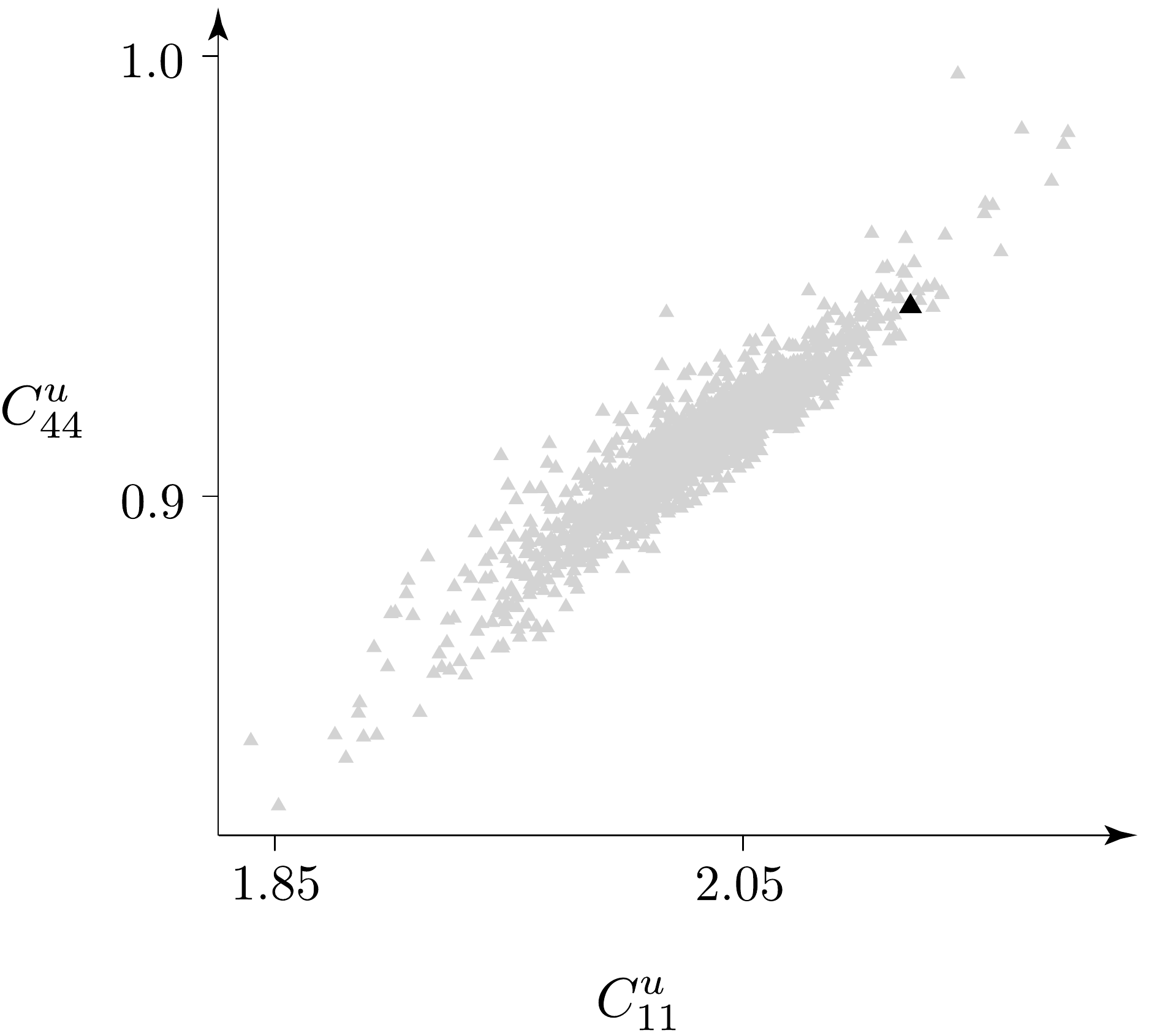}\hspace{0.3in}
\includegraphics[scale=0.42]{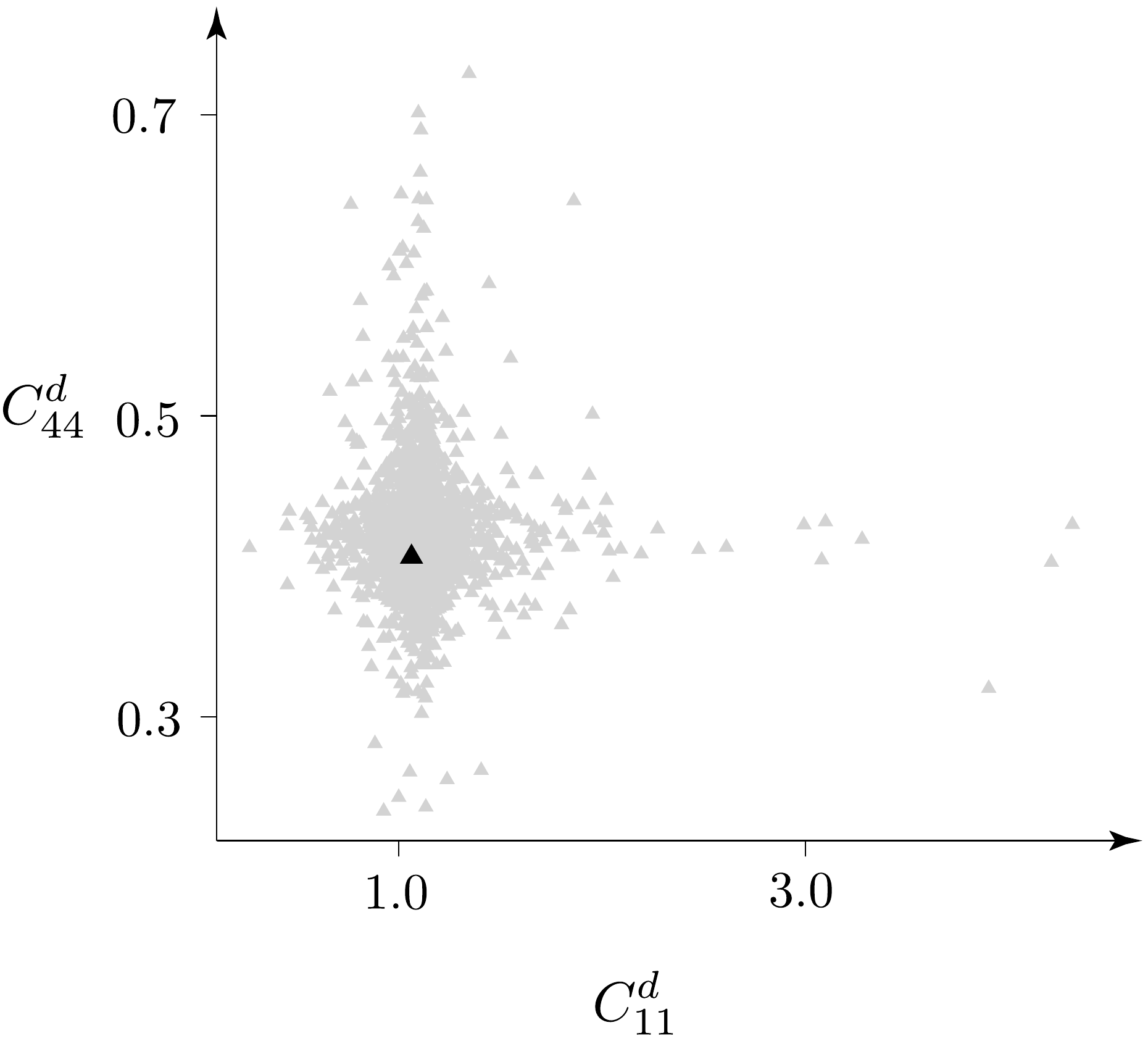}
}
\caption{Relations between elasticity parameters for the input error of $+5\%$}
\label{fig:corr_+5}
\end{figure}
\begin{figure}
\centerline{
\hspace*{-0.1in}
\includegraphics[scale=0.42]{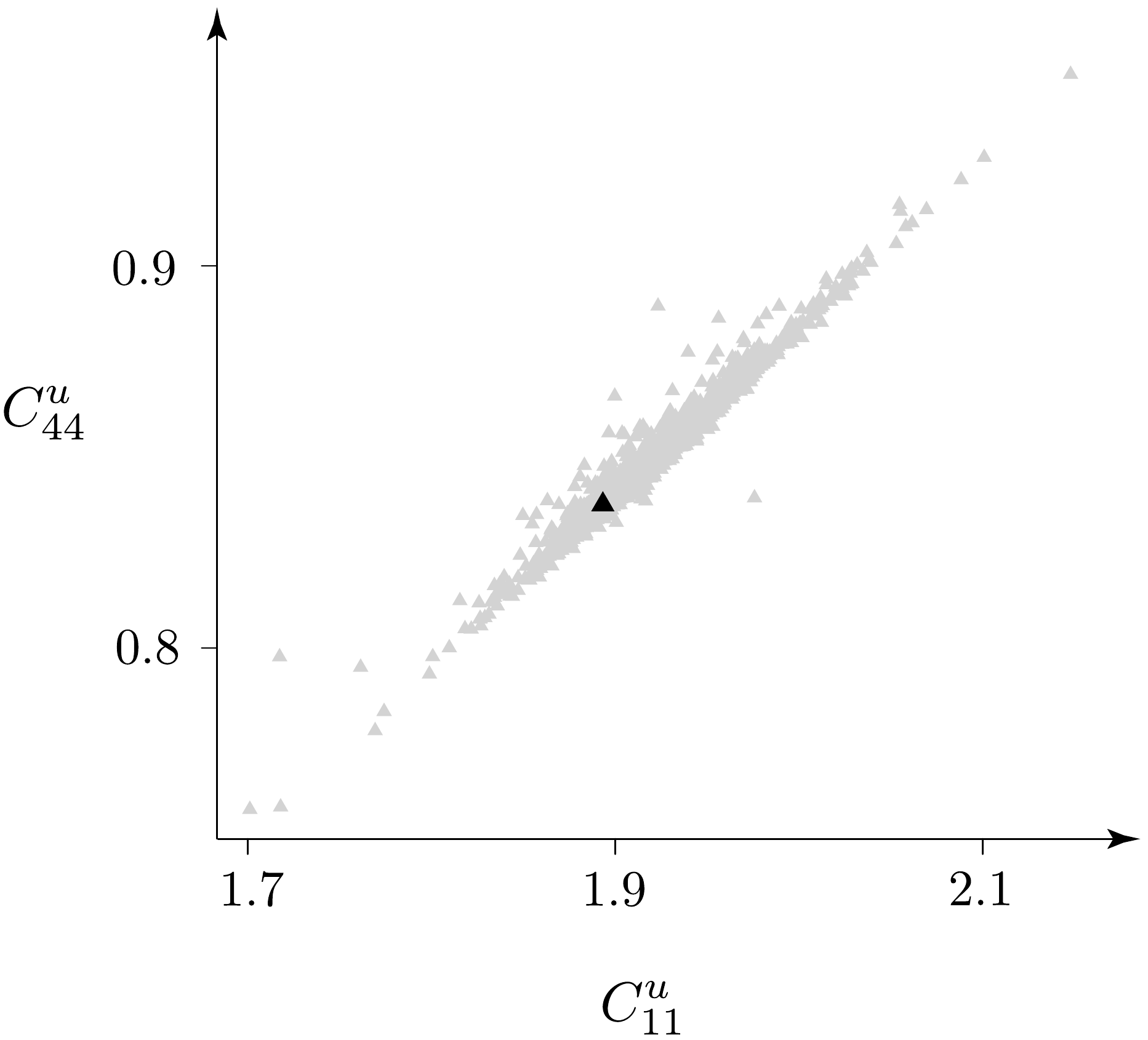}\hspace{0.3in}
\includegraphics[scale=0.42]{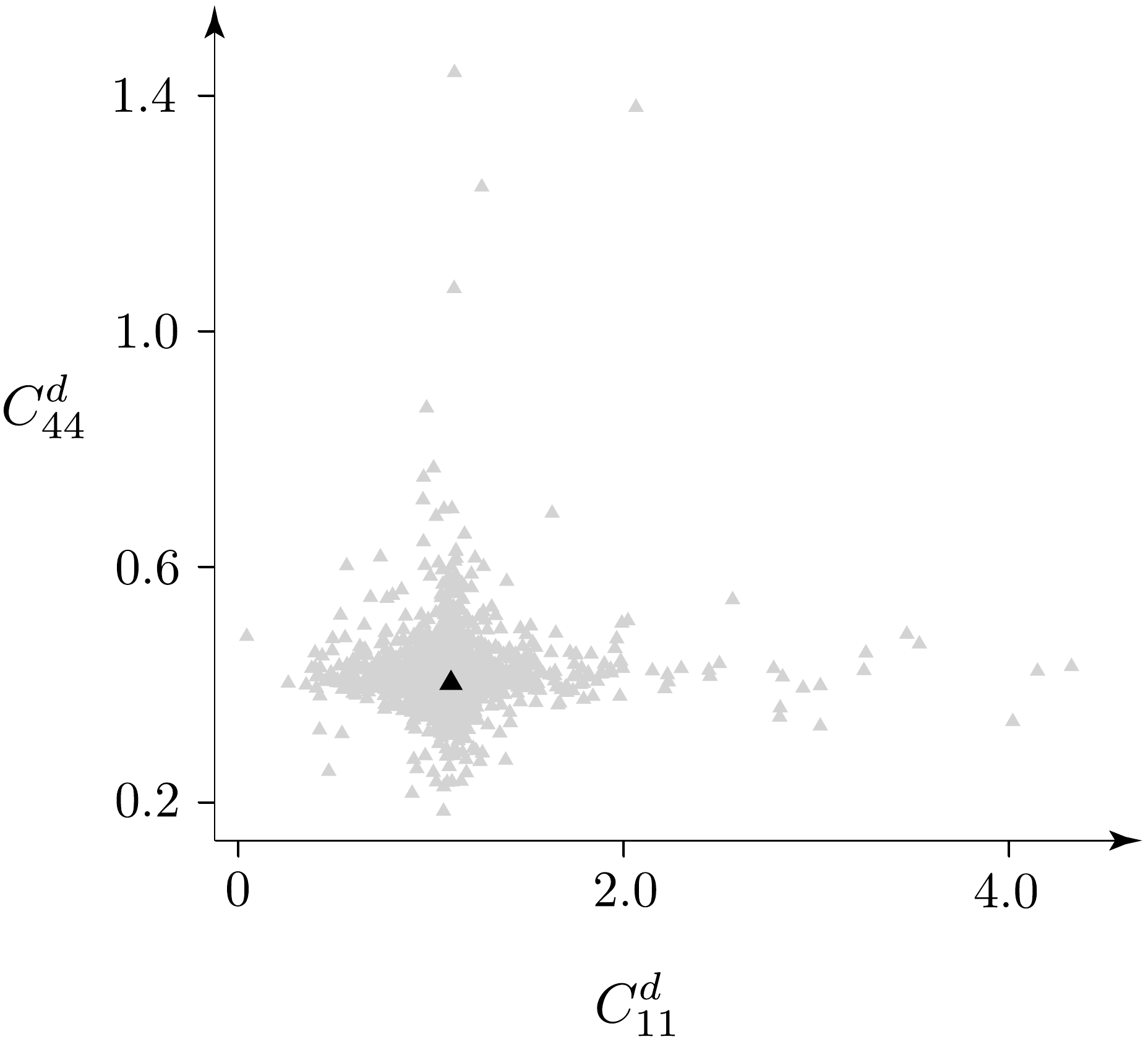}
}
\caption{Relations between elasticity parameters for for the input error of $-5\%$}
\label{fig:corr_-5}
\end{figure}
To gain an insight into the effects of input errors on the inverse process, we examine correlations between the inverted values of $C_{11}$ and $C_{44}$\,, in the layer and in the halfspace, for input with and without errors.
In Figures~\ref{fig:corr_unperturbed}--\ref{fig:corr_-5}, the left-hand plot corresponds to the layer and the right-hand plot to the halfspace.
For each case, we consider only the fundamental mode and $\omega=15~{\rm s}^{-1}$\,.

In Figure~\ref{fig:corr_unperturbed}, we illustrate the correlation for input without errors.
In Figures~\ref{fig:corr_+5} and \ref{fig:corr_-5}, we illustrate the correlation for the speed perturbed by $+5\%$ and $-5\%$\,, respectively.
The values along the axes for the left-hand and right-hand plots are to be scaled as follows.
Figure~\ref{fig:corr_unperturbed}: $\times 10^9$\,, $\times 10^{11}$\,; Figure~\ref{fig:corr_+5}: $\times 10^{10}$\,, $\times 10^{11}$\,; Figure~\ref{fig:corr_-5}:  $\times 10^{10}$\,, $\times 10^{11}$\,.

The three left-hand plots, which refer to the layers, exhibit a linear relation between $C_{11}$ and $C_{44}$\,.
In each plot, the black triangle corresponds to the optimal Pareto solution.
In units of $10^{10}~{\rm N}/{\rm m}^2$, they are as follows.
Figure~\ref{fig:corr_unperturbed}:  $C_{11}^u=1.999$\,, $C_{44}^u=0.893$\,, $C_{11}^d=10.405$\,, $C_{44}^d=3.919$\,, as in Table~\ref{tab:solutions_II};
Figure~\ref{fig:corr_+5}: $C_{11}^u=2.121$\,, $C_{44}^u=0.943$\,, $C_{11}^d=10.736$\,, $C_{44}^d=4.061$\,,  as in Table~\ref{tab:solutions_+5per};
Figure~\ref{fig:corr_-5}: $C_{11}^u=1.894$\,, $C_{44}^u=0.838$\,, $C_{11}^d=11.066$\,, $C_{44}^d=4.042$\,, which do not appear in any table.

Let us consider a linear regression,
\begin{equation*}
C^u_{44}=0.43\,C^u_{11} + 2.97\times 10^8\,,	
\end{equation*}
\begin{equation*}
C^u_{44}=0.37\,C^u_{11} + 1.63\times 10^9\,,
\end{equation*}
\begin{equation*}
C^u_{44}=0.42\,C^u_{11} + 3.96\times 10^8\,,	
\end{equation*}
for each left-hand plot, respectively.
The slope is similar for each case, the intercept varies slightly more.
These results show that the ratio of elasticity parameters is preserved along the Pareto front.

The linear relation between $C^u_{11}$ and $C^u_{44}$ might be due to the asymptotic behaviour of the fundamental mode, which---for both quasi-Rayleigh waves and Love waves---depends on the values of~$C_{44}^u$\,.
Hence, the value of $C_{11}^u$ has to adjust itself, in accordance with solutions along the Pareto front.
A detailed explanation of this phenomenon requires further study.
 
The three right-hand plots, which refer to the halfspace, exhibit neither the linear relation nor a significant shift of the optimal Pareto solution, marked by a black triangle.
These results suggest that the layer elasticity parameters are more sensitive to the input errors than are the halfspace parameters, which is consistent with results listed in  Table~\ref{tab:solutions_per_5}, which is similar to Table~\ref{tab:solutions_+5per}, except values are expressed in percentages, for the first mode at $\omega=15~{\rm s}^{-1}$\,, where the layer parameters are further, in a percentage sense, from their actual values than are the halfspace parameters.
\subsection{Sensitivity}
Dalton et al.~\cite{dalton} (Section~4.2) conclude that the fundamental mode is mainly sensitive to the layer parameters; higher modes are sensitive to both the layer and halfspace properties.
Examining the values for $C_{44}^u$ and $C_{44}^d$\,, in Table~\ref{tab:solutions_II_per}, it appears that---for the fundamental mode---the inverse results for $C_{44}^u$ are more accurate than for $C_{44}^d$\,; for higher modes, the discrepancy in the accuracy becomes less pronounced and in certain cases it is even reversed.
For the second mode, the results for $C_{11}^d$ and $C_{44}^d$ are more accurate for the lower frequency, which is $15~{\rm s}^{-1}$\,, than for $60~{\rm s}^{-1}$ and  $100~{\rm s}^{-1}$\,, which is in agreement with Dalton et al.~\cite{dalton}.

$C_{11}^u$ and $C_{11}^d$ do not appear in the target function for Love waves, so they are not constrained by the minimization of the target function for these waves, but only by the minimization of the target function for quasi-Rayleigh waves.
This can be exemplified by a statistical analysis.
For instance, for the fundamental mode at $\omega=15~{\rm s}^{-1}$\,, the standard deviations of $C_{11}^d$\,, $C_{44}^d$\,, $C_{11}^u$ and $C_{44}^u$ are $1.51\times10^{10}$\,, $3.19\times 10^9$\,, $5.21\times 10^8$ and $2.25\times 10^8$\,, respectively. 

The distributions of $C_{11}^u$ and $C_{11}^d$ are more varied than those of $C_{44}^u$ and $C_{44}^d$\,.
$C_{11}^u$ and $C_{11}^d$ seem to be less constrained and are characterized by more outliers, which suggest an effect of the absence of $C_{11}^d$ and $C_{11}^u$ in the Love-wave forward solver.
Also, in general, the distribution of inverse results in the halfspace is more varied and exhibits more outliers than the distribution in the layer.
These conclusions are drawn from an analysis of unfiltered results for data without errors; filtering would remove values beyond $\pm 50\%$ of the true value.
\section{Conclusion}
\label{sec:gwdiscussion}
The obtained results support the use of the Love-wave and quasi-Rayleigh-wave speeds, which are measurable on the surface, in a Pareto Joint Inversion.
Also, the Particle Swarm Optimization is an efficient method for finding the sought minima.
 
For error-free input, the inferred model parameters are accurate, without any further constraints.
For input with errors, these parameters become significantly less accurate, which indicates the error-sensitivity of the process.
Also, correlations between the inverted elasticity parameters indicate that the layer parameters are more sensitive to input errors than the halfspace parameters.
Further constraints, such as the information about the layer mass density or its elasticity parameters, might be necessary for a reliable inference of model parameters.

In agreement with Dalton et al.~\cite{dalton}, the fundamental mode is more sensitive to the layer parameters.
Also, in agreement with Dalton et al.~\cite{dalton}, higher modes are sensitive to both the layer and halfspace properties; for the second mode, the inverse results are more accurate for low frequencies.

Solutions presented in this paper are obtained by considering a single mode at a time.
 A method that could use several modes simultaneously might be a significant improvement of efficiency.
\section*{Acknowledgments}
We wish to acknowledge discussions with Piotr Stachura and Theodore Stanoev, as well as the graphic support of Elena Patarini.
This research was performed in the context of  The Geomechanics Project supported by Husky Energy. 
Also, this research was partially supported by the Natural Sciences and Engineering Research Council of Canada, grant 238416-2013, and by the Polish National Science Center under contract No.\ DEC-2013/11/B/ST10/0472.
\bibliographystyle{unsrt}
\bibliography{GWinverse}
\end{document}